\documentclass[10pt,prd,superscriptaddress,preprintnumbers,amsmath,amssymb,nofootinbib,twocolumn]{revtex4-2}
\usepackage{xcolor}
\usepackage{graphicx}
\usepackage{tabularx}
\usepackage{latexsym}
\usepackage{float}
\usepackage{epsfig}
\usepackage{subfigure}
\usepackage{multirow}

\RequirePackage{flushend}
\RequirePackage[colorlinks,citecolor=blue,urlcolor=blue,linkcolor=blue]{hyperref}
\newcommand{\greencheck}{{\color{green}\checkmark}} % Define a green check mark command
\newcommand{\blackcross}{{\color{black}\texttimes}}
\usepackage{verbatim}
\begin{abstract}
We extend the Standard Model with the general effective Hamiltonian for the quark level transition $c \to s \ell \nu$ with a complete set of four fermion operators including right-handed neutrinos. The current experimental measurements in charm decays are compatible with the Standard Model predictions and are used to constrain the new physics. With the available experimental data, we fit a $\chi^2$ function to get the best-fit values of the NP WCs. We investigate the impact of allowed new physics in the observables such as differential branching fraction, forward-backward asymmetry, lepton polarization asymmetry, and convexity parameter in the semileptonic decay $B_c^+ \to B_s \mu^+ \nu_{\mu}$. The different types of new physics scenarios significantly affect these considered observables. Future experimental information on these observables can help disentangle the structure of new physics.
\end{abstract}
\begin{document}
\title{New Physics effects with right-handed neutrinos in semileptonic decay $B_c^+ \to B_s \mu^+ \nu_{\mu}$ }
%%%%%%%%%%%%%%%%%%%%%%%%%%%%%%%%%%%%%%%%%%%%%%%%%%%%%%%%%%%%%%%
\author{Priyanka Boora}
\email{2020rpy9601@mnit.ac.in}
\affiliation{Department of Physics, Malaviya National Institute of Technology Jaipur, India}
%%%%%%%%%%%%%%%%%%%%%%%%%%%%%%%%%%%%%%%%%%%%%%%%%%%%%%%%%%%%%%%%
\author{Dinesh Kumar}
\email{dineshsuman09@gmail.com}
\affiliation{Department of Physics, University of Rajasthan, Jaipur 302004, India}
%%%%%%%%%%%%%%%%%%%%%%%%%%%%%%%%%%%%%%%%%%%%%%%%%%%%%%%%%%%%%%%
\author{Kavita Lalwani}
\email{kavita.phy@mnit.ac.in}
\affiliation{Department of Physics, Malaviya National Institute of Technology Jaipur, India}
%%%%%%%%%%%%%%%%%%%%%%%%%%%%%%%%%%%%%%%%%%%%%%%%%%%%%%%%%%%%%%%
\maketitle
\section{Introduction}  
The search for new physics (NP) beyond the Standard Model (SM) has been an integral part of research in particle physics. The discrepancy between the experimental measurement and SM prediction can hint at new physics beyond SM. No new particle beyond SM has yet been observed at the experimental facilities to support the physics beyond the SM. The world average of lepton flavor universality observable, defined by $\mathcal{R}_{D^{(*)}} = \frac{\mathcal{R}(B \to D^* \tau \bar{\nu)}}{\mathcal{R}(B \to D^* \ell \bar{\nu)}}$, shows a deviation from SM prediction at the level of $3.31 \sigma$ \cite{hflav}. The branching ratio $\mathcal{R}_{J/\Psi} = \frac{\mathcal{R}(B_c \to J/\psi \tau \bar{\nu)}}{\mathcal{R}(B_c \to J/\psi \mu \bar{\nu)}} = 0.71 \pm 0.17 \pm 0.18$ \cite{LHCb:2017vlu} measured by the LHCb collaboration is in 1.8 $\sigma$ tension with SM predictions from different works \cite{Anisimov:1998uk, Ivanov:2006ni, Hernandez:2006gt, Huang:2007kb, Wang:2008xt, Wang:2012lrc, Watanabe:2017mip, Issadykov:2018myx, Tran:2018kuv, Hu:2018veh, Wang:2018duy, Hu:2019qcn, Leljak:2019eyw, Azizi:2019aaf}. These measurements along with others provide a hint for new physics.

Headway in the direction of searching for new physics with bottom-charmed mesons offers a good platform through three-body semileptonic decays. As with B decays, the charm mesons have a good probability of transition \cite{khlopov1978effects, gershtein1976leptonic} through semileptonic or leptonic decays and can be a key tool to explore the new physics. For such a study, we choose the bottom-charm meson $B_{c}^{+}$, which was first observed via the decay mode $ B_{c}^{\pm} \to J/\psi \ell^{\pm} \nu$ by the collider detector at Fermilab (CDF collaboration) \cite{CDF:1998ihx}. Studies of $B_c^+$ with the Mesogenesis and Baryogenesis mechanisms \cite{Elahi:2021jia, Elor:2020tkc} for the branching fraction and CP-violation which relies on generating the observed baryon asymmetry and dark matter make this meson more interesting to study with exclusive semileptonic decays. This particle decays through weak interactions, and decay can occur through charm and beauty quark transitions.

The decay of the $B_c^+$ meson through the charm quark has a dominant contribution over the full decay width with smaller available phase space \cite{colangelo1993qcd, Beneke:1996xe, Anisimov:1998uk, Kiselev:2000pp}. In our study, we focus on the semileptonic decay $ B_{c}^{+} \to B_{s} \bar{\ell} \nu_{\ell}$ induced by the Cabibbo-favored $c \to s \ell \nu_{\ell}$ quark level transition by considering the new physics in the second generation of leptons. We consider our new physics study with right-handed neutrino in an effective field theory approach.

In order to circumvent the existing phenomenological restrictions on the EFT operators containing left-handed neutrino (LHN) fields, light right-handed neutrinos (RHNs) have been proposed \cite{Beltran:2022ast, Ligeti:2016npd, Asadi:2018wea, Greljo:2018ogz, Robinson:2018gza, Azatov:2018kzb, Babu:2018vrl, Heeck:2018ntp, Asadi:2018sym, Bardhan:2019ljo, Shi:2019gxi, Gomez:2019xfw}. Sterile neutrinos are singlets with properties that are not coherent with the involved charged electroweak companions, and we consider the massless right-handed neutrinos to exclude the cosmological and astrophysical limits. With massless neutrino, two different neutrino chiralities will not interfere in the decay probabilities \cite{Mandal:2020htr}.

The previous phenomenological study on $ B_{c}^{\pm} \to B_{s,d}^{(*)} \bar{\ell} \nu_{\ell}$ with left-handed neutrinos gives the hint of the deviations from standard Model \cite{Colangelo:2021dnv} in the differential branching fraction observable and introduction of right-handed neutrinos in the B-decays for a model-independent study gives a presence of new physics to resolve the anomalies in semileptonic $B$- decays \cite{Mandal:2020htr, Penalva:2021wye, Datta:2022czw}. These two studies motivate us to scratch the $B_c^+ \to B_s \mu^+ \nu_{\mu}$ decay in the presence of a right-handed neutrino. In ref. \cite{Colangelo:2021dnv}, authors analyzed differential branching fraction of $B_c^+ \to B_s \mu^+ \nu_{\mu}$ decay and its correlation with the branching fraction of $B_c^+ \to B_s^* \mu^+ \nu_{\mu}$ decay.

In this paper, we aim to explore the structure of new physics in the charm sector based on dimension-six operators of effective field theory including RHNs along with LHNs. Following the refs. \cite{Mandal:2020htr, Penalva:2021wye, Datta:2022czw}, we introduced the RHNs in the leptonic current of the decay and rewrite the Hamiltonian of ref. \cite{Colangelo:2021dnv} by taking into account the charge-conjugated lepton fields, since the charge-conjugated fields do the same role as the original field for the particles \cite{Alvarado:2024lpq}. We consider the NP scenarios introduced in ref. \cite{Mandal:2020htr} based on the nature of the mediator particle. In this, we analyze all the possible observables for a pseudoscalar three-body semileptonic decay i.e., differential branching fraction, forward-backward asymmetry, lepton polarization asymmetry, and convexity parameter. The new physics Wilson coefficients (WCs) are constrained by using the available relevant experimental measurements in the charm sector and the best-fit values are obtained by using the $\chi^2$ minimization. 

The present work is organized as follows. In section \ref{TF} the theoretical framework for our analysis is defined. Section \ref{npc} defines the constraint for the new physics by following the numerical analysis for the best-fit values of Wilson coefficients in Section \ref{NA}. In section \ref{observables} observables for the analysis are defined with the required form factor information. In section, \ref{result} results of our numerical fits are discussed and the new physics effects on observables in $B_c^+ \to B_s \mu^+ \nu_{\mu}$ decay are described. Section \ref{summary} summarizing the sensitivity of all observables in different scenarios. Finally, in section \ref{conclusion} we provide the conclusion of our analysis. All technical information, such as hadronic amplitudes, form factors, and fit parameters are compiled in appendices \ref{app-A} and \ref{app-B}.

\section{Theoretical Framework}
\label{TF}
We work in an effective field theory approach using the framework of Ref. \cite{Mandal:2020htr}. We considered the most general effective Hamiltonian by considering the new physics with the left-handed and right-handed neutrino operators.

\subsection{Effective Hamiltonian}
\label{EH}
We extend the Standard Model by considering the low-energy effective Hamiltonian comprising the full set of dimension-six operators, which govern the quark level general form as transition $c \to s \ell \nu$, including left-handed and right-handed neutrinos in
\begin{equation}
\label{equ-1}
\begin{split}
H_{eff} = \frac{4 G_F V_{cs}}{\sqrt{2}}\Big[(1 + C_{LL}^V) \mathcal{O}_{LL}^V +  C_{RL}^V \mathcal{O}_{RL}^V + C_{LR}^V  \mathcal{O}_{LR}^V \\ + C_{RR}^V \mathcal{O}_{RR}^V 
+ C_{LL}^S \mathcal{O}_{LL}^S + C_{RL}^S \mathcal{O}_{RL}^S  
+ C_{LR}^S \mathcal{O}_{LR}^S + \\ C_{RR}^S \mathcal{O}_{RR}^S + C_{LL}^T \mathcal{O}_{LL}^T + C_{RR}^T \mathcal{O}_{RR}^T\Big] + h.c.,
\end{split}
\end{equation}

where $C_{XY}^i$ are the new physics Wilson coefficients and $\mathcal{O}_{XY}^i$, with $i=(V,S,T)$ and $X,Y = (L,R)$, are the four fermion operators which are defined as:
\begin{eqnarray}
\label{equ-2}
\begin{aligned}
&\mathcal{O}_{L(R)L}^V = (\bar{s} \gamma^{\alpha}P_{L(R)} c) (\bar{\nu_{\mu}} \gamma_{\alpha}P_{L} \mu) \\
&\mathcal{O}_{L(R)R}^V = (\bar{s} \gamma^{\alpha}P_{L(R)} c) (\bar{\nu_{\mu}} \gamma_{\alpha}P_{R} \mu)\\
&\mathcal{O}_{L(R)L}^S = (\bar{s} P_{L(R)} c) (\bar{\nu_{\mu}} P_L \mu)  \\
&\mathcal{O}_{L(R)R}^S = (\bar{s} P_{L(R)} c) (\bar{\nu_{\mu}} P_{R} \mu) \\
&\mathcal{O}_{L(R)L(R)}^T = \delta_{ij}(\bar{s} \sigma^{\alpha\beta} P_{L(R)} c) (\bar{\nu_{\mu}} \sigma_{\alpha\beta} P_{L(R)} \mu) \\
\end{aligned}
\end{eqnarray}

where the quark flavors are defined in the mass basis, $V_{cs}$ is the CKM-matrix element, $G_F$ is the fermi constant and $P_{L(R)} = \frac{(1 \pm \gamma_5)}{2}$ are projection operators.
The inclusion of RHNs in effective Hamiltonian only transfigures the leptonic currents without any changes in the hadronic current therefore, the same form factors can be used in the decay amplitude as used for the case of LHN operators in the ref. \cite{Colangelo:2021dnv}. The generalized Hamiltonian, as in Eq. (\ref{equ-1}), has been studied for the $b \to c$ transitions in connection with the anomalies in semileptonic $B \to D \tau \bar{\nu}$, $B \to D^{*} \tau \bar{\nu}$ and $\Lambda_{b} \to \Lambda_{c} \tau \bar{\nu_{\tau}}$ decays to check their sensitivity for the new physics \cite{Mandal:2020htr, Penalva:2021wye}.  
%%%%%%%%%%%%%%%%%%%%%%%%%%%%%%%%%%%%%%%%%%%%%%%%%%%%%%%%%%%%%%%%%%%%%%

\section{New Physics Constraints}
\label{npc}
The experimental observations in $c \to s \ell \nu$ that are currently available can constrain the new physics. We take into account the observations from $c \to s \ell \nu$ governed leptonic and semileptonic decays.

\subsection{Leptonic decay: $D_{s}^{+} \to \mu^{+} \nu_{\mu}$}
The branching fraction of leptonic decay $D_{s}^{+} \to \mu^{+} \nu_{\mu}$ receives the contribution from the NP operators in the presence of right-handed neutrino and the expression for this can be given by:
\begin{equation}
\label{equ-13}
\begin{aligned}
& {\mathcal{B}(D_s^+ \to \mu^+ \nu_{\mu})} = \\ &
\frac{G_{F}^2}{8\pi} |V_{cs}|^{2} f_{D_{s}^{+}}^{2} m_{D_{s}^{+}} m_{\mu}^{2} \Big(1 - \frac{m_{\mu}^{2}}{m_{D_{s}^{+}}^{2}}\Big)^{2} \tau_{D_{s}^{+}} \\ & 
\Big[ \Big|(1 + C_{LL}^{V} - C_{RL}^{V}) + \frac{m_{D_{s}^{+}}^{2}}{m_{\mu}(m_{c}+m_{s})} (C_{RL}^{S} - C_{LL}^{S})\Big|^{2}  \\ &
+ \Big|(C_{RR}^{V} - C_{LR}^{V}) + \frac{m_{D_{s}^{+}}^{2}}{m_{\mu}(m_{c}+m_{s})} (C_{LR}^{S} - C_{RR}^{S})\Big|^{2}\Big] 
\end{aligned}		
\end{equation}

The hadronic dynamics of this leptonic decay is encoded in the single parameter decay constant $f_{D_{s}^{+}} = (248.0 \pm 1.6)$ MeV which is calculated by using lattice QCD and we used the value determined by the FLAG working group \cite{aoki2022flag}.
The experimental measurement of this branching ratio is given in Table \ref{tab-1}.
\vspace{-7mm}
\subsection{Semileptonic decays}
The semileptonic decays also get the NP contribution with right-handed neutrinos. These semileptonic decays can be very useful in constraining short-distance NP coefficients. The semileptonic decays are also governed by the same quark level transition $c \to s\ell \nu$ like leptonic decay hence these will provide further constraining power with leptonic decays.
\subsubsection{$D \to P \mu^{+} \nu_{\mu}$}
We consider the semileptonic decays $D \to P \mu^{+} \nu_{\mu}$, where D are the $D^{0}$, $D^{+}$ mesons and P denotes the pseudoscalar $K^{-}$ and $\Bar{K^{0}}$ mesons respectively\cite{Fleischer:2019wlx}. The differential branching fraction, including right-handed neutrinos, is given by

\begin{equation}
\begin{aligned}
&\frac{d \mathcal{B}}{d q^2}(D \rightarrow P \mu^{+} \nu_{\mu}) = \\ &\frac{G_F^2 V_{c s}^2}{192 m_{D}^3 \pi^3} q^2 \sqrt{\lambda_{P}(q^2)}\left(1-\frac{m_\mu^2}{q^2}\right)^2 \tau_{D} \\
& \times \left\{\left(\left|1+C_{L L}^V+C_{R L}^V\right|^2+\left|C_{L R}^V+C_{R R}^V\right|^2\right) \right.\\
& \left[\left(H_{V, 0}^P\right)^2\left(\frac{m_\mu^2}{2 q^2}+1\right)+\frac{3 m_\mu^2}{2 q^2}\left(H_{V, s}^P\right)^2\right] \\
& +\frac{3}{2}\left(H_{S,s}^P\right)^2\left(\left|C_{R L}^S+C_{L L}^S\right|^2+\left|C_{R R}^S+C_{L R}^S\right|^2\right) \\
& +8\left(\left|C_{L L}^T\right|^2+\left|C_{R R}^T\right|^2\right)\left(H_T^P\right)^2\left(1+\frac{2 m_\mu^2}{q^2}\right) \\
& +3 Re\left[\left(1+C_{L L}^V+C_{R L}^V\right)\left(C_{R L}^S+C_{L L}^S\right)^* \right. \\
& \left. +\left(C_{L R}^V+C_{R R}^V\right)\left(C_{R R}^S+C_{L R}^S\right)^*\right] \frac{m_\mu}{\sqrt{q^2}} H_{S,s}^P H_{V, s}^P \\
& \left.-12 Re\left[\left(1+C_{LL}^V+C_{R L}^V\right) C_{ LL}^{T*} 
+ (C_{R R}^V+C_{L R}^V\right) C_{R R}^{T *}\right] \\
& \left. \frac{m_\mu}{\sqrt{q^2}} H_T^P H_{V, 0}^P\right\} 
\end{aligned}
\end{equation}

where H's are the helicity amplitudes written in terms of the form factors $f_0(q^2), f_{+}(q^2), f_T(q^2)$ and form factors are calculated using z-series parameterization \cite{Lubicz:2017syv, Lubicz:2018rfs} given in \ref{app-B}. The experimental measurements for semileptonic decays $D \to P \mu^+ \nu_{\mu}$ are listed in Table \ref{tab-1}.

\subsubsection{$D \to V \mu^{+} \nu_{\mu}$}
We continue to constrain the new physics operators further by considering semileptonic $D \to V$ decays, where D are $D^{0}$, $D^{+}$ and $D_{s}^{+}$ mesons and V are the $K^{*-}$, $\bar{K}^{*0}$ and $\phi$ mesons \cite{Fleischer:2019wlx, Leng:2020fei}. The differential branching fraction for this semileptonic decay with right-handed neutrinos can be given by:\\
\begin{equation}
\begin{aligned}
&\frac{d \mathcal{B}}{d q^2}(D \rightarrow V \mu^{+} \nu_{\mu}) = \\ & \frac{G_F^2 V_{c s}^2}{192 m_{D}^3 \pi^3} q^2 \sqrt{\lambda_{V}(q^2)}\left(1-\frac{m_\mu^2}{q^2}\right)^2 \tau_{D} \\
& \times \left\{\left(\left|1+C_{L L}^V\right|^2+\left|C_{R L}^V\right|^2+\left|C_{L R}^V\right|^2+\left|C_{R R}^V\right|^2\right) \right. \\
& \left[\left((H_{V, +}^V)^2 + (H_{V, -}^V)^2 + (H_{V, 0}^V)^2\right)\left(\frac{m_\mu^2}{2 q^2}+1\right) +\frac{3 m_\mu^2}{2 q^2}\left(H_{V, t}^V\right)^2\right] \\
& - 2 Re\left[\left(1+C_{L L}^V\right) C_{RL}^{V*}+C_{L R}^V C_{RR}^{V*}\right] \\
&\left[\left((H_{V, 0}^V)^2 + 2H_{V, +}^V H_{V, -}^V\right)\left(\frac{m_\mu^2}{2 q^2}+1\right)+\frac{3 m_\mu^2}{2 q^2}\left(H_{V, t}^V\right)^2\right] \\
& +\frac{3}{2}(H_{S}^V)^2\left(\left|C_{R L}^S-C_{L L}^S\right|^2+\left|C_{R R}^S-C_{L R}^S\right|^2\right) \\
&+8\left(\left|C_{L L}^T\right|^2+\left|C_{R R}^T\right|^2\right) 
\left((H_{T,+}^V)^2 + (H_{T,-}^V)^2 + (H_{T,0}^V)^2\right) \\
&\left(1+\frac{2 m_\mu^2}{q^2}\right) 
+3 Re\left[\left(1+C_{L L}^V-C_{R L}^V\right)\left(C_{R L}^S-C_{L L}^S\right)^* \right. \\
&\left.+\left(C_{L R}^V-C_{R R}^V\right)\left(C_{R R}^S-C_{L R}^S\right)^*\right] 
\frac{m_\mu}{\sqrt{q^2}} H_{S}^V H_{V, t}^V \\
& \left.
-12 Re\left[\left(1+C_{LL}^V\right) C_{LL}^{T *}+C_{R R}^V C_{R R}^{T *}\right] \right. \\
& \left.\frac{m_\mu}{\sqrt{q^2}} \left(H_{T,0}^V H_{V, 0}^V + H_{T,+}^V H_{V, +}^V - H_{T,-}^V H_{V, -}^V\right)\right.\\
& \left.+12 Re\left[C_{RL}^V C_{LL}^{T *}+C_{L R}^V C_{R R}^{T *}\right] \right. \\
& \left.\frac{m_\mu}{\sqrt{q^2}} \left(H_{T,0}^V H_{V, 0}^V + H_{T,+}^V H_{V, -}^V - H_{T,-}^V H_{V, +}^V\right)\right \}
\end{aligned}
\end{equation}

where H's are the hadronic helicities and form factors used from the calculation result of LCSR \cite{Wu:2006rd, bowler1995improved} given in \ref{app-B}, the measurement for the semileptonic decay $D \to V \ell \nu$ is given in Table \ref{tab-1}.
%%%%%%%%%%%%%%%%%%%%%%%%%%%%%%%%%%%%%%%%%%%%%%%%%%%%%%%%%%%%%%%%%%%%%%%%%

\section{Numerical Analysis}
\label{NA}
The analysis of the new physics effects requires a set of input values of Wilson coefficients $C_{XY}^i$ defined in Eq (\ref{equ-1}). The experimental measurements of leptonic and semileptonic decay of $D$ mesons are considered to constrain the allowed NP.

\begin{table}
	\begin{center}
		\caption{Branching ratios of the required leptonic and semi-leptonic decays available experimental measurement used in numerical analysis.}
		\setlength{\tabcolsep}{13pt}
		\label{tab-1}    
		\begin{tabular}{ll}
			\hline\noalign{\smallskip}
			Mode & Experimental Value   \\
			\noalign{\smallskip}\hline\noalign{\smallskip}
			$D_{s}^{+} \to \mu^{+} \nu_{\mu}$ & (5.49 $\pm$ 0.16) $\times 10^{-3}$ \cite{particle2020review} \\ 
			$D^{0} \to K^{*-} \mu^{+} \nu_{\mu}$ & (1.89 $\pm$ 0.24) $\times 10^{-2}$ \cite{particle2020review} \\ 
			$D^{0} \to K^{-} \mu^{+} \nu_{\mu}$ & (3.41 $\pm$ 0.04) $\times 10^{-2}$  \cite{particle2020review} \\
			$D^{+} \to \Bar{K^{*0}} \mu^{+} \nu_{\mu}$ & (5.27 $\pm$ 0.15) $\times 10^{-2}$ \cite{particle2020review} \\
			$D^{+} \to \Bar{K^{0}} \mu^{+} \nu_{\mu}$ &(8.76 $\pm$ 0.19) $\times 10^{-2}$ \cite{particle2020review} \\
			$D_{s}^{+} \to \phi \mu^{+} \nu_{\mu}$ & (1.90 $\pm$ 0.50) $\times 10^{-2}$  \cite{particle2020review} \\
			\noalign{\smallskip}\hline
		\end{tabular}
	\end{center}
\end{table}
We adopt the methodology of the $\chi^2$ fitting to get the best-fit values of the new physics Wilson coefficients. The $\chi^2$ as a function of the Wilson coefficient (WCs) is written as $\chi^2(C^i) = \frac{\Big(\mathcal{O}_{th}(C^i) - \mathcal{O}_{exp}\Big)^2}{\sigma_i^2}$, 
Where $\mathcal{O}_{th}(C^i)$ is the theoretical expression in terms of NP Wilson coefficients and $\mathcal{O}_{exp}$ is the experimental value of the corresponding observable. The $\sigma_i$ are the uncertainties by adding the theoretical and experimental errors in the quadrature. For a particular new physics scenario, the most likely value of the corresponding new physics operators is obtained by minimization through the MINUIT package \cite{james1975minuit, james1998minuit}.

%%%%%%%%%%%%%%%%%%%%%%%%%%%%%%%%%%%%%%%%%%%%%%%%%%%%%%%%%%%%%%%%%%%%%%%%%%%%%
\section{NP effects on observables in $B^{+}_{c} \to B_{s} \mu^+ \nu_{\mu}$ }
\label{observables}
First of all, we write the differential decay distribution of the $B^{+}_{c} \to B_{s} \mu^+ \nu_{\mu}$ which will receive the NP contribution in the presence of right-handed neutrinos. The expression for the two-fold differential decay distribution can be written as
\begin{equation}
\label{equ-3}
\begin{aligned}
&\frac{d^2 \Gamma(B^{+}_c \rightarrow B_s \mu^{+} \nu_{\mu})}{d q^2 d \cos \theta_\mu} = \\
& \mathcal{N} \left\{\mathcal{L}_0(q^2) + \mathcal{L}_1(q^2) \cos \theta_\mu + \mathcal{L}_2(q^2) (\cos \theta_\mu)^2 \right\}
\end{aligned}
\end{equation}

where $ \mathcal{N} = \frac{G_F^2 V_{c s}^2 q^2 \sqrt{\lambda_{B_{s}}(q^2)}}{256 m_{B_{c}}^3 \pi^3}\left(1-\frac{m_\mu^2}{q^2}\right)^2$ and $q^2 = (p_{\mu^{+}} + p_{\nu})^2$  is the squared momentum transferred to the muon-neutrino pair, $\theta_{\mu}$  is the polar angle of the muon momentum in the rest frame of the $\mu^{+} \nu$ pair, w.r.to the z-axis defined by the $B_c^+$ meson momentum in the rest frame of $B_s$ frame, and $\lambda_{B_s}$ is the triangle function or shorthand notation of the K{\"a}llen function $\lambda_{B_{s}}\left(q^2\right) \equiv \lambda\left(m_{B_{c}}^2, m_{B_{s}}^2, q^2\right) = ((m_{B_{c}} - m_{B_{s}})^2 - q^2) ((m_{B_{c}} + m_{B_{s}})^2 - q^2)$.

The angular coefficients are given as 
\begin{equation}
\label{equ-5}
\small\begin{aligned}
& \mathcal{L}_0\left(q^2\right)=\left|\mathcal{W}_0^L-\frac{2 m_\mu}{\sqrt{q^2}} \mathcal{W}_T^L\right|^2+\frac{m_\mu^2}{q^2}\left|\mathcal{W}_t^L+\frac{\sqrt{q^2}}{m_\mu} \mathcal{W}_S^L\right|^2 \\ 
& +(L \leftrightarrow R), \\
& \mathcal{L}_1\left(q^2\right)=\frac{2 m_\mu^2}{q^2} Re \left[\left(\mathcal{W}_0^L-\frac{2 \sqrt{q^2}}{m_\mu} \mathcal{W}_T^L\right)\left(\mathcal{W}_t^{L *}+\frac{\sqrt{q^2}}{m_\mu} \mathcal{W}_S^{L *}\right)\right] \\ 
&+(L \leftrightarrow R), \\
& \mathcal{L}_2\left(q^2\right)=-\left(1-\frac{m_\mu^2}{q^2}\right)\left(\left|\mathcal{W}_0^L\right|^2-4\left|\mathcal{W}_T^L\right|^2\right)+(L \leftrightarrow R)
\end{aligned}
\end{equation}

Where
\begin{equation}
\label{equ-6}
\begin{aligned}
\mathcal{W}_0^L & =\left(1+C_{L L}^V+C_{R L}^V\right) H_{V, 0}^s, &  \; \mathcal{W}_0^R &=\left(C_{L R}^V+C_{R R}^V\right) H_{V, 0}^s, \\
\mathcal{W}_t^L & =\left(1+C_{L L}^V+C_{R L}^V\right) H_{V, t}^s, & \; \mathcal{W}_t^R &=\left(C_{L R}^V+C_{R R}^V\right) H_{V, t}^s, \\
\mathcal{W}_S^L & =\left(C_{R L}^S+C_{L L}^S\right) H_S^s, & \; \mathcal{W}_S^R &=\left(C_{R R}^S+C_{L R}^S\right) H_S^s, \\
\mathcal{W}_T^L & =2 C_{L L}^T H_T^s, & \;
\mathcal{W}_T^R &=2 C_{R R}^T H_T^s
\end{aligned}
\end{equation}

The H's are the Hadronic helicity amplitudes, and their explicit expressions are given in \ref{app-A}.

Integrating over the angle $\theta_{\mu}$, the $q^{2}$ distribution of the $B^{+}_{c} \to B_{s} \mu^+ \nu_{\mu}$ governed  by the Eq (\ref{equ-1}) reads
\begin{equation}
\label{equ-8}
\begin{aligned}
&\frac{d \varGamma}{d q^2}(B^{+}_c \rightarrow B_s \mu^{+} \nu_{\mu}) = \\
& \frac{G_F^2 V_{c s}^2}{192 m_{B_{c}}^3 \pi^3} q^2 \lambda_{B_{s}}^{1 / 2}\left(q^2\right)\left(1-\frac{m_\mu^2}{q^2}\right)^2 \\
& \times \left\{\left(\left|1+C_{L L}^V+C_{R L}^V\right|^2+\left|C_{L R}^V+C_{R R}^V\right|^2\right) \right. \\
&\left[\left(H_{V, 0}^s\right)^2\left(\frac{m_\mu^2}{2 q^2}+1\right)+\frac{3 m_\mu^2}{2 q^2}\left(H_{V, t}^s\right)^2\right] \\ 
& +\frac{3}{2}\left(H_S^s\right)^2 
\left(\left|C_{R L}^S+C_{L L}^S\right|^2+\left|C_{R R}^S+C_{L R}^S\right|^2\right) \\
& + 8\left(\left|C_{L L}^T\right|^2+\left|C_{R R}^T\right|^2\right)\left(H_T^s\right)^2\left(1+\frac{2 m_\mu^2}{q^2}\right) \\
& +3 Re\left[\left(1+C_{L L}^V+C_{R L}^V\right)\left(C_{R L}^S+C_{L L}^S\right)^* \right. \\
&\left. +\left(C_{L R}^V+C_{R R}^V\right)\left(C_{R R}^S+C_{L R}^S\right)^*\right] \frac{m_\mu}{\sqrt{q^2}} H_S^s H_{V, t}^s \\
& \left. -12 Re\left[\left(1+C_{LL}^V+C_{R L}^V\right) C_{ LL}^{T *} + \left(C_{R R}^V+C_{L R}^V\right) C_{R R}^{T *}\right] \right.\\
&\left. \frac{m_\mu}{\sqrt{q^2}} H_T^s H_{V, 0}^s\right\} 
\end{aligned}
\end{equation}

The expression for forward-backward asymmetry in terms of the NP Wilson coefficients can be given as 
\begin{equation}
\label{equ-9}
\begin{aligned}
&\mathcal{A}_{FB} = \\
&\frac{1}{\frac{d\Gamma}{dq^{2}}} \big[ \int_{0}^{1} - \int_{-1}^{0} \big] d cos\theta_\mu \frac{d^2 \Gamma}{d q^2 d \cos \theta_\mu} =
\frac{\mathcal{L}_1(q^2)}{2 \mathcal{L}_0(q^2) + \frac{2}{3} \mathcal{L}_2(q^2)}
\end{aligned}
\end{equation}

We also define the Lepton polarization asymmetry \cite{Becirevic:2019tpx} and convexity \cite{Becirevic:2020rzi} as
\begin{equation}
\label{equ-10}
\mathcal{A}_{\lambda_{\mu}} = \frac{\frac{d\Gamma_{\lambda_{\mu} = -1/2}}{dq^{2}} - \frac{d\Gamma_{\lambda_{\mu} = +1/2}}{dq^{2}} }{\frac{d\Gamma}{dq^{2}}} = 1 - 2 \Big(\frac{\frac{d\Gamma_{\lambda_{\mu} = 1/2}}{dq^{2}}}{\frac{d\Gamma}{dq^{2}}} \Big) 
\end{equation}

\begin{equation}
\label{equ-11}
\begin{aligned}
&\mathcal{A}_{\pi/3} =
\frac{1}{\frac{d\Gamma}{dq^{2}}} \big[ \int_{1/2}^{1} - \int_{-1/2}^{1/2} + \int_{-1}^{-1/2} \big] d cos\theta_\mu \frac{d^2 \Gamma}{d q^2 d \cos \theta_\mu} = \\
& \frac{\mathcal{L}_2(q^2)}{4 \mathcal{L}_0(q^2) + \frac{4}{3} \mathcal{L}_2(q^2)}
\end{aligned}
\end{equation}
For the analysis of the decay $B^{+}_c \rightarrow B_s \mu^+ \nu_{\mu}$ decay its form factors $f_0$ and $f_{+}$ are computed from the lattice QCD \cite{Cooper:2020wnj}. These are evaluated over the full range of $q^2$; $ m_{\mu}^2 < q^2 < (m_{B_{c}^+} - m_{B_{s}})^2$ obtained from the chain fit of the results of highly improved staggered quark (HISQ) method and non-relativistic QCD. 

The form factors $f_0$, $f_+$ are expressed by the truncated power series of $z_p$ in the physical continuum limit with a constraint of $f_{0}(0) = f_{+}(0)$ as
\begin{equation}
\label{equ-12}
\begin{aligned}
& f(q^{2})=P(q^{2}) \sum_n^N A_n z_p(q^{2})^n
\end{aligned}
\end{equation} 
The semileptonic region $ m_{\mu}^2 < q^2 < (m_{B_{c}^+} - m_{B_{s}})^2$ is mapped within the unit circle 
$z(q^{2})=\frac{\sqrt{t_{+}-q^{2}}-\sqrt{t}_{+}}{\sqrt{t_{+}-q^{2}}+\sqrt{t}_{+}}$ with 
$t_{+}=\left(m_{B_c}+m_{B_{s}}\right)^2$ on the real axis region. For calculation optimization, $z$ is rescaled by 
$z_p\left(q^2\right)=\frac{z\left(q^2\right)}{\left|z\left(M_{res}^2\right)\right|}$ 
with the mass $M_{res}$ of the nearest $c\bar{s}$ meson pole $M_{D_{s}^{*}}$ = 2.112 GeV \cite{particle2020review}.  The factor  
$P(q^{2}) = \left(1-q^2 / M_{\mathrm{res}}^2\right)^{-1}$ represents the pole structure of the form factors and correlation error is included with the data set of covariance matrix $A_{n}$. The fit parameter values are given in Table \ref{tab-2}.

\begin{table}%[H]
	\begin{center}
		\caption{Input parameters in our analysis are taken from Ref\cite{particle2020review}.}
		\setlength{\tabcolsep}{15pt}
		\label{tab-2}       % Give a unique label
		\begin{tabular}{ll}
			\hline\noalign{\smallskip}
			Parameter & Numerical Value   \\
			\noalign{\smallskip}\hline\noalign{\smallskip}
			$V_{cs}$ & 0.987 $\pm$ 0.011 \\ 
			$m_{B_c}$ & 6274.9 $\pm$ 0.8 MeV \\
			$m_{B_s}$ & 5366.88 $\pm$ 0.14 MeV \\
			$m_{D_s^+}$ & 1968.34 $\pm$ 0.07 MeV \\
			$m_{D^{0}}$ & 1864.83 $\pm$ 0.05 MeV \\
			$m_{D^{+}}$ & 1869.65 $\pm$ 0.05 MeV \\
			$m_{K^{0}}$ & 497.611 $\pm$ 0.013 MeV \\
			$m_{K^{*0}}$ & 895.55 $\pm$ 0.20 MeV \\
			$m_{K^{-}}$ & 493.677 $\pm$ 0.016 MeV \\
			$m_{K^{*-}}$ & 895.5 $\pm$ 0.8 MeV \\
			$m_{\phi}$ & 1019.461 $\pm$ 0.016 MeV \\
			$m_{c}$ & 1.27 $\pm$ 0.02 GeV \\
			$m_{s}$ & 0.093 $\pm$ 0.011 GeV \\ 
			$m_{\mu}$ & 105.658 MeV \\
			$\tau_{B_{c}}$ & (0.510 $\pm$ 0.009) $\times 10^{-12}$ sec \\
			$\tau_{D_{s}^+}$ & (5.04 $\pm$ 0.04) $\times 10^{-13}$ sec\\
			$\tau_{D^0}$ & (4.101 $\pm$ 0.015) $\times 10^{-13}$ sec \\
			$\tau_{D^+}$ & (1.104 $\pm$ 0.007) $\times 10^{-12}$ sec\\
			\noalign{\smallskip}\hline
		\end{tabular}
	\end{center}
\end{table}

%%%%%%%%%%%%%%%%%%%%%%%%%%%%%%%%%%%%%%%%%%%%%%%%%%%%%%%%%%%%%%%%%%%%%%%%
\section{Results}
\label{result}
We consider the new physics effects in a model-independent way in the framework of EFT where we included the additional operators with right-handed neutrinos. The experimental measurements are used to constrain the NP WCs. The effects of NP WCs is explored in the observables defined in section \ref{observables} in semileptonic decay $B_c^+ \to B_s \mu^+ \nu_{\mu}$.

The possible high-scale NP mediators are identified in ref. \cite{Mandal:2020htr} which can generate new physics operators involving right-handed neutrinos. The single new physics mediators can be integrated out which can give one or more new physics operators contributing to $c \to s \ell \nu$ processes. We used these scenarios for our analysis and listed in Table \ref{tab-3}. 
Scenarios 1 and 2 arise when we take into account the contribution of right-handed neutrino with and without SM-like contributions, respectively. In scenarios, 4, 5, and 7, the $'a'$ scenario is generated with the contribution from RHN operators while the $'b'$ scenario contains the contribution also from the LHN operators. Scenarios 3, 6, and 8 do not generate any left-handed operators. 
For all the defined scenarios in Table \ref{tab-3}, we analyze the $q^2$ spectrum of the differential branching fraction, forward-backward asymmetry, lepton polarization asymmetry, and convexity parameter. We provide the $q^2$ spectrum for these observables for SM as well as for the central values of the NP WCs.

\begin{table}%[H]
	\begin{center}
		\caption{List of the scenarios with quantum numbers (Q.No.) and involved new physics operators.}
		\label{tab-3}       % Give a unique label
		\begin{tabular}{lll}
			\hline\noalign{\smallskip}
			Scenarios & Q. No. & Operators   \\
			\noalign{\smallskip}\hline\noalign{\smallskip}
			Scenario 1 \\ (RHN + SM-like) & - & $\mathcal{O}_{LL}^{V}, \mathcal{O}_{LR}^{V}, \mathcal{O}_{RR}^{V}$, \\ &&$\mathcal{O}_{LR}^{S}, \mathcal{O}_{RR}^{S}, \mathcal{O}_{RR}^{T}$  \\
			\hline
			Scenario 2 \\ (RHN) & - &$\mathcal{O}_{LR}^{V}, \mathcal{O}_{RR}^{V}, \mathcal{O}_{LR}^{S}$,\\ &&$\mathcal{O}_{RR}^{S}, \mathcal{O}_{RR}^{T}$\\
			\hline
			Scenario 3 \\ (Vector Boson)  & $V^{\mu}$ (1,1,-1) & $\mathcal{O}_{RR}^{V}$\\
			\hline
			Scenario 4 \\ (Scalar Boson)  & $\varphi$ (1,2,1/2) &  (a) $\mathcal{O}_{LR}^{S}, \mathcal{O}_{RR}^{S}$ \\
			&&(b) $\mathcal{O}_{LL}^{S}, \mathcal{O}_{RL}^{S}$, \\
			&& $\mathcal{O}_{LR}^{S}, \mathcal{O}_{RR}^{S}$ \\
			\hline
			Scenario 5 \\ (Leptoquark) & $U_{1}^{\mu}$ (3,1,2/3) & (a) $\mathcal{O}_{RR}^{V},\mathcal{O}_{LR}^{S}$\\
			&& (b) $\mathcal{O}_{LL}^{V}, \mathcal{O}_{RL}^{S}$, \\
			&& $\mathcal{O}_{RR}^{V}, \mathcal{O}_{LR}^{S}$ \\
			\hline
			Scenario 6 \\ (Leptoquark) & $\tilde{R}_{2}$ (3,2,1/6) & $\mathcal{O}_{RR}^{S}, \mathcal{O}_{RR}^{T}$ \\ 
			\hline
			&& (a) $\mathcal{O}_{RR}^{V}, \mathcal{O}_{RR}^{S}, \mathcal{O}_{RR}^{T}$ \\      
			Scenario 7 \\ (Leptoquark) & $S_{1}$ ($\bar{3}$,1,1/3) & (b) $\mathcal{O}_{LL}^{V}, \mathcal{O}_{LL}^{S}, \mathcal{O}_{LL}^{T}$, \\ 
			&& $\mathcal{O}_{RR}^{V}, \mathcal{O}_{RR}^{S}, \mathcal{O}_{RR}^{T}$ \\
			\hline
			Scenario 8 \\ (Leptoquark) & $\tilde{V}_{2}^{\mu}$ ($\bar{3}$,2,-1/6) & $\mathcal{O}_{LR}^{S}$ \\
			\noalign{\smallskip}\hline
		\end{tabular}
	\end{center}
\end{table}

\subsection{Scenario 1}
In this scenario, the contribution from RHN operators and SM-like operator $C_{LL}^V$ is considered, and we get the following fit-values and correlation for NP WCs as given in Table \ref{tab-sc1}.

\begin{table} [H]
	\caption{Scenario 1: Fit values and correlation}
	\vspace{3pt}
	\setlength{\tabcolsep}{3pt}
	\label{tab-sc1}
	\renewcommand{\arraystretch}{1.5}
	\centering\begin{tabular}{lc|llllll}
		\hline
		\multirow{2}{*}{WCs}&\multirow{2}{*}{Fit Values}&\multicolumn{6}{c}{Correlation} \\        
		\cline{3-8}		
		&&$C_{LL}^V$&$C_{LR}^V$ & $C_{RR}^V$ & $C_{LR}^S$&$C_{RR}^S$&$C_{RR}^T$\\
		\hline
		$C_{LL}^V$ &$-0.066 \pm 0.141$ &1 &0.9  &0.7 &0&0.1 &0.1 \\
		
		$C_{LR}^V$ &$-0.367 \pm 0.380$&0.9 &1  &0.6 & 0 & 0 &-0.2 \\
		
		$C_{RR}^V$ &$ 0.116 \pm 0.119$ &0.7 &0.6 &1 &-0.3&0 &0.5 \\
		
		$C_{LR}^S$ &$-0.097 \pm 0.017$  &0 &0 &-0.3 &1&0.9 &-0.2 \\
		
		$C_{RR}^S$&$-0.094 \pm 0.017$ &0.1 &0  &0 &0.9&1 &0.2 \\		
		
		$C_{RR}^T$&$ 0.017 \pm 0.332$ &0.1 &-0.2 &0.5 &-0.2&0.2&1 \\
		\hline
	\end{tabular}
\end{table}

\begin{figure}%[H]
	\centering
	\includegraphics[width=0.22\textwidth]{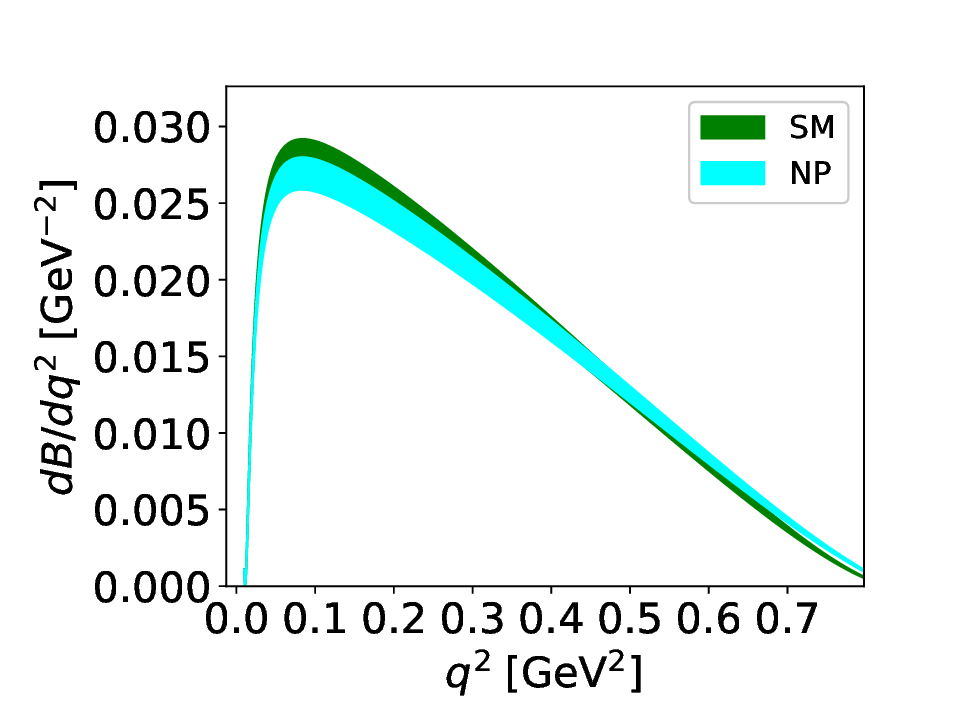}
	\includegraphics[width=0.22\textwidth]{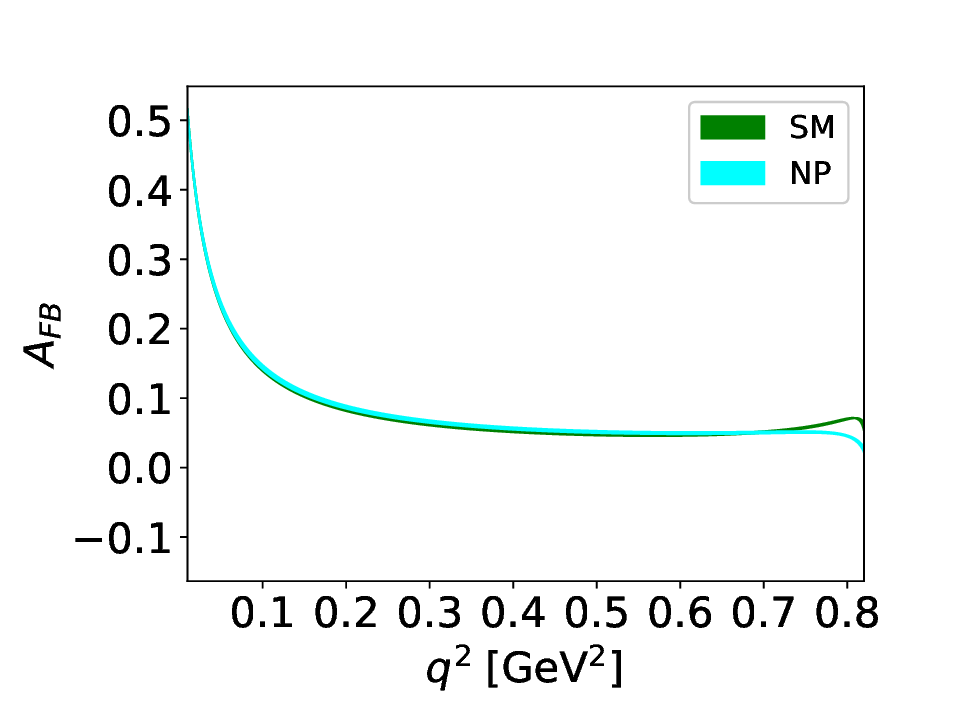}
	\includegraphics[width=0.22\textwidth]{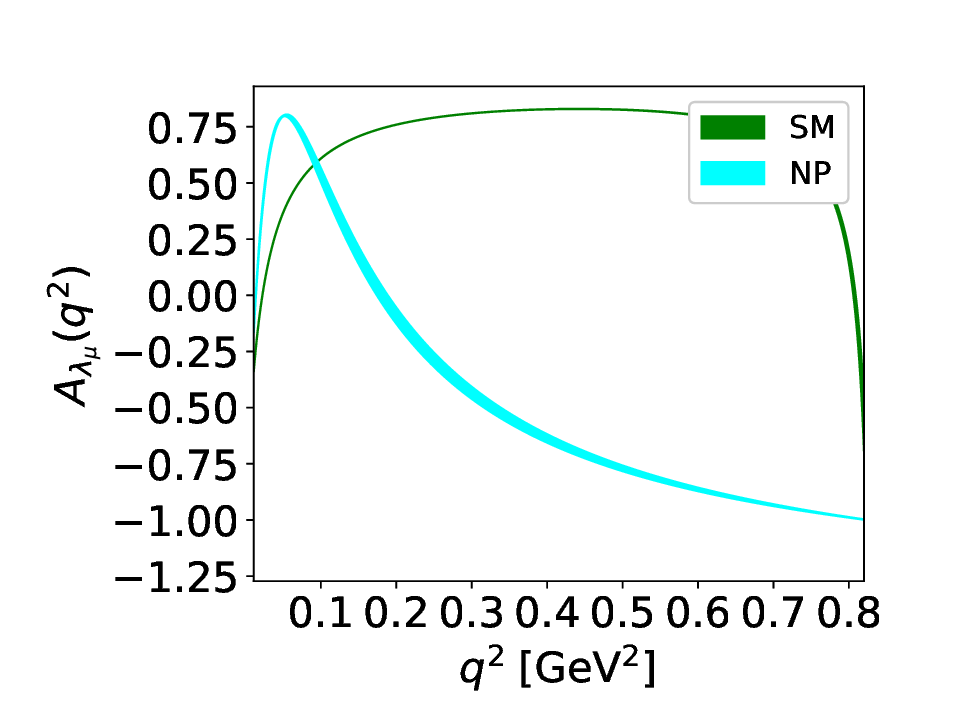}
	\includegraphics[width=0.22\textwidth]{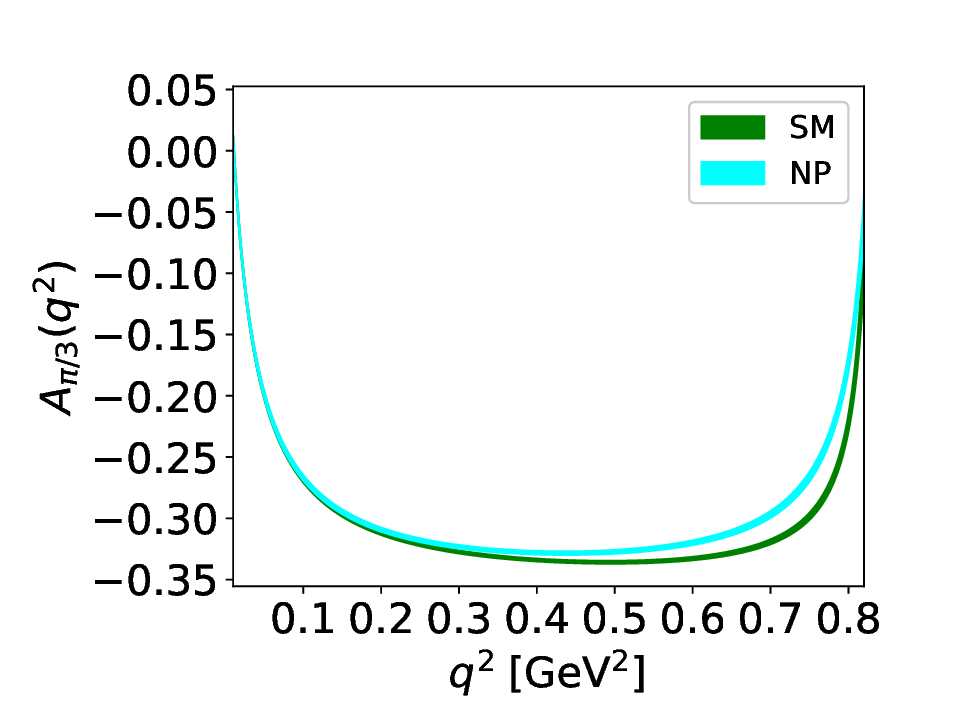}
	\caption{$q^2$ spectra of scenario 1 for the differential branching fraction $\frac{d\mathcal{B}}{dq^2}$ (top left), forward-backward asymmetry $A_{FB}$ (top right), lepton polarization asymmetry $A_{\lambda_{\mu}}$ (bottom left), and convexity $A_{\pi/3}$ (bottom right). The Standard Model (SM green band) and new physics (NP cyan band) both include uncertainty from the form factors.}
	\label{fig-3}       % Give a unique label
\end{figure}

We give the predictions for all four observables in SM as well as NP by taking the central values of NP WCs which are shown in figure \ref{fig-3}. The lepton polarization asymmetry is the most sensitive observable with a zero crossing symmetry at $q^2 \sim 0.18$ $GeV^2$ and on the other hand NP convexity is deviating from SM for $q^2 > 0.35$ $ GeV^2$ values. However, the differential branching fraction and forward-backward asymmetry do not show any significant deviation from the SM.
\subsection{Scenario 2}
In this general scenario, the contribution to $B_c^+ \to B_s \mu^+ \nu_{\mu}$ receives only from the presence of all RHN operators $\big($$\mathcal{O}_{LR}^V$, $\mathcal{O}_{RR}^V$, $\mathcal{O}_{LR}^S$, $\mathcal{O}_{RR}^S$, $\mathcal{O}_{RR}^T$$\big)$ in the theory. This is also a general model-independent assumption as no new type of mediator is considered. The fit-values with correlation are following in Table \ref{tab-sc2}:

\begin{table} [H]
	\caption{Scenario 2: Fit values and correlation}
	\setlength{\tabcolsep}{3pt}
	\vspace{2pt}
	\label{tab-sc2}
	\renewcommand{\arraystretch}{1.5}
	\centering\begin{tabular}{lc|lllll}
		\hline
		\multirow{2}{*}{WCs}&\multirow{2}{*}{Fit Values}&\multicolumn{5}{c}{Correlation} \\        
		\cline{3-7}		
		&&$C_{LR}^V$ & $C_{RR}^V$ & $C_{LR}^S$&$C_{RR}^S$&$C_{RR}^T$\\
		\hline	
		$C_{LR}^V$ &$-0.193 \pm 0.207$& 1 & 0.4 & 0.1 & 0.01 & -0.5 \\
		
	$C_{RR}^V$ &$ 0.113 \pm 0.184$ &0.4 & 1 & 0.01 & 0.1 & 0.3 \\
		
	$C_{LR}^S$ &$ 0.008 \pm 0.044$  &0.1 & 0.01 & 1 & 0.8 & -0.1 \\
		
		$C_{RR}^S$&$ 0.013 \pm 0.044$ & 0.01 & 0.1 & 0.8 & 1 & 0.1 \\		
		
		$C_{RR}^T$&$ 0.010 \pm 0.349$ & -0.5 & 0.3 & -0.1 & 0.1& 1 \\
		\hline
	\end{tabular}
\end{table}

\begin{figure}%[H]
	\centering
	\includegraphics[width=0.22\textwidth]{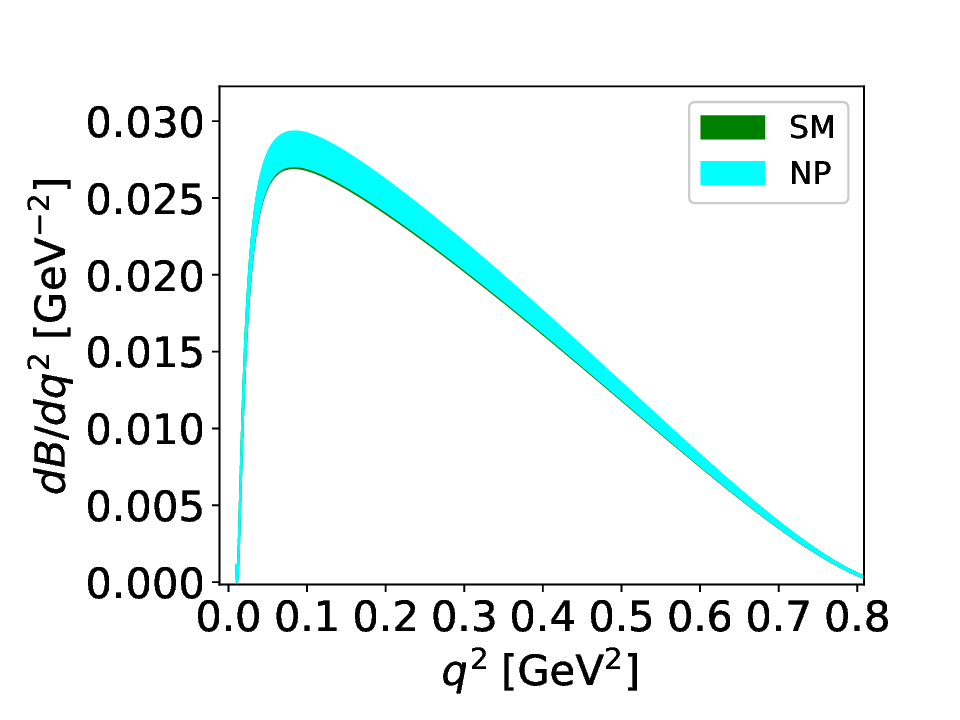}
	\includegraphics[width=0.22\textwidth]{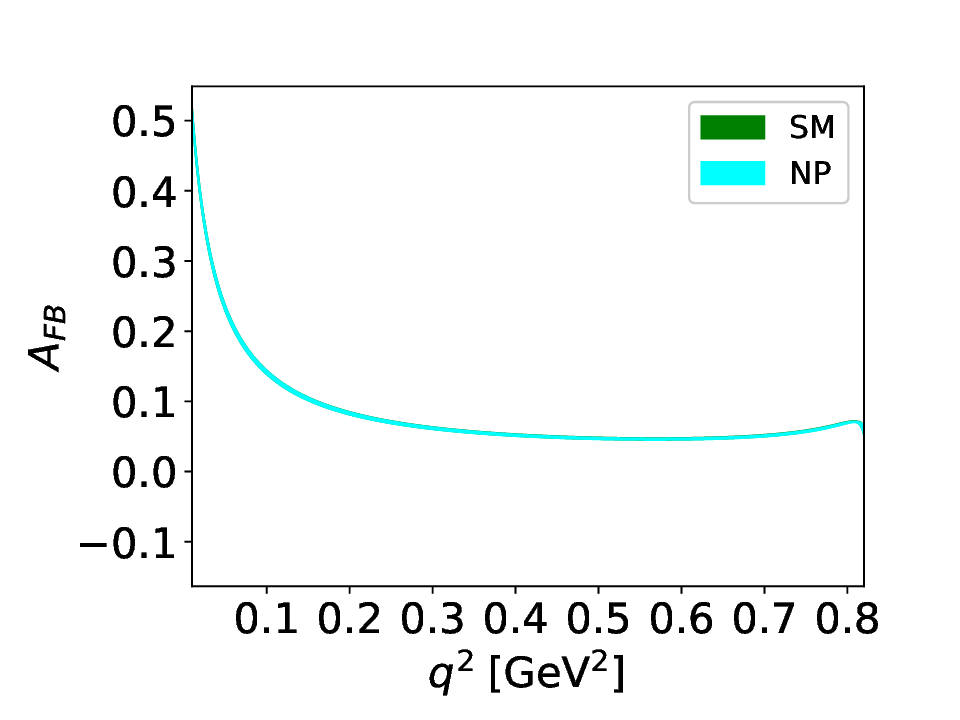}
	\includegraphics[width=0.22\textwidth]{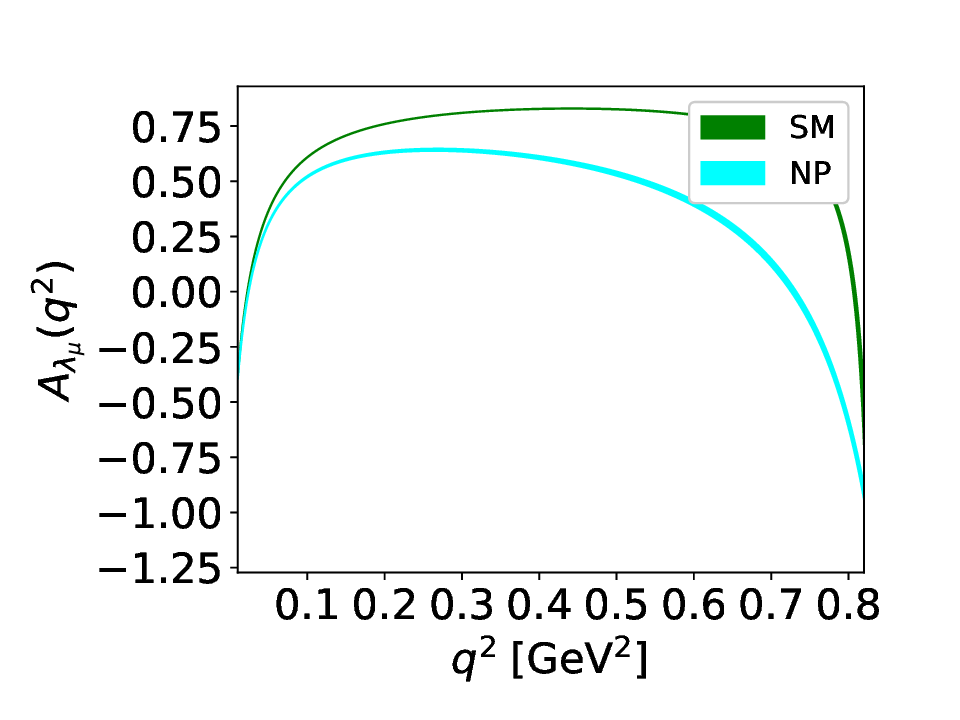}
	\includegraphics[width=0.22\textwidth]{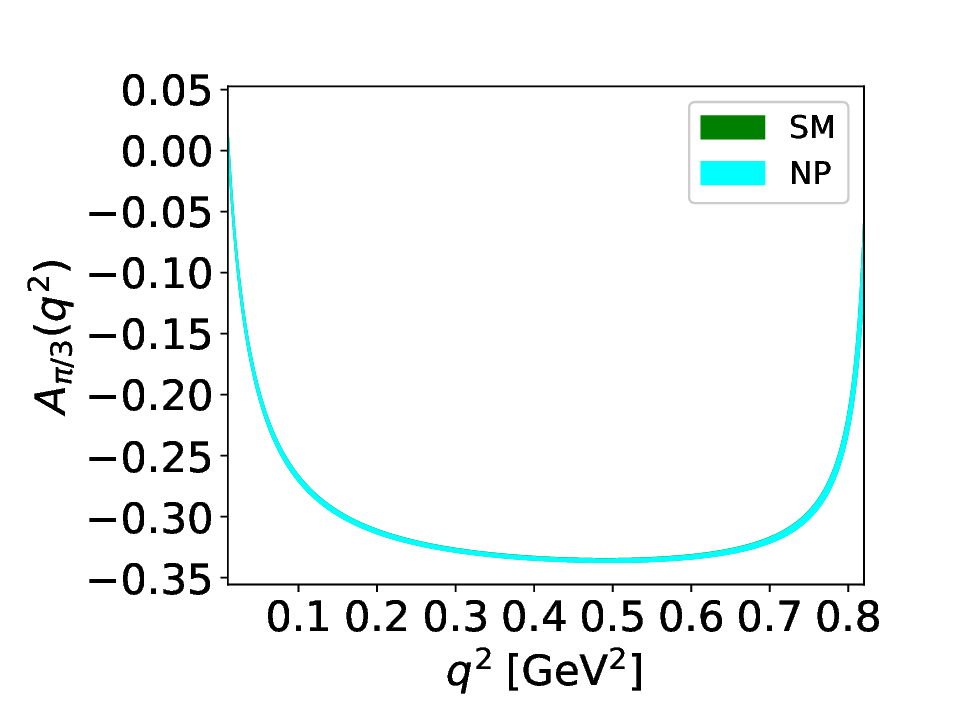}
	\caption{$q^2$ spectra of scenario 2 for the differential branching fraction $\frac{d\mathcal{B}}{dq^2}$ (top left), forward-backward asymmetry $A_{FB}$ (top right), lepton polarization asymmetry $A_{\lambda_{\mu}}$ (bottom left), and convexity $A_{\pi/3}$ (bottom right). The Standard Model (SM green band) and new physics (NP cyan band) both include uncertainty from the form factors.}
	\label{fig-4}       % Give a unique label
\end{figure}

	The prediction plots for the observables in semileptonic decay $B_c^+ \to B_s \mu^+ \nu_{\mu}$ for scenario 2 are shown in figure \ref{fig-4}. When only the right-handed neutrinos are involved from the new physics side then the lepton polarization asymmetry shows some sensitivity by deviating from the SM band with zero crossing at higher $q^2 \sim 0.7$ $GeV^2$ side, on the other side for differential branching fraction, forward-backward asymmetry and convexity the NP band almost overlaps with SM band.

\subsection{Scenario 3}
The new physics mediator vector boson $V^{\mu}(1,1,-1)$ has only vector interaction with RHN so there is only one new physics operator $\mathcal{O}^V_{RR}$ is present in this scenario. The  fit value for this operator is:

	\begin{eqnarray}
	\label{equ-sc3fv}
	C_{RR}^V = -0.166 \pm 0.077 
	\end{eqnarray}

\begin{figure}%[H]
	\centering
	\includegraphics[width=0.22\textwidth]{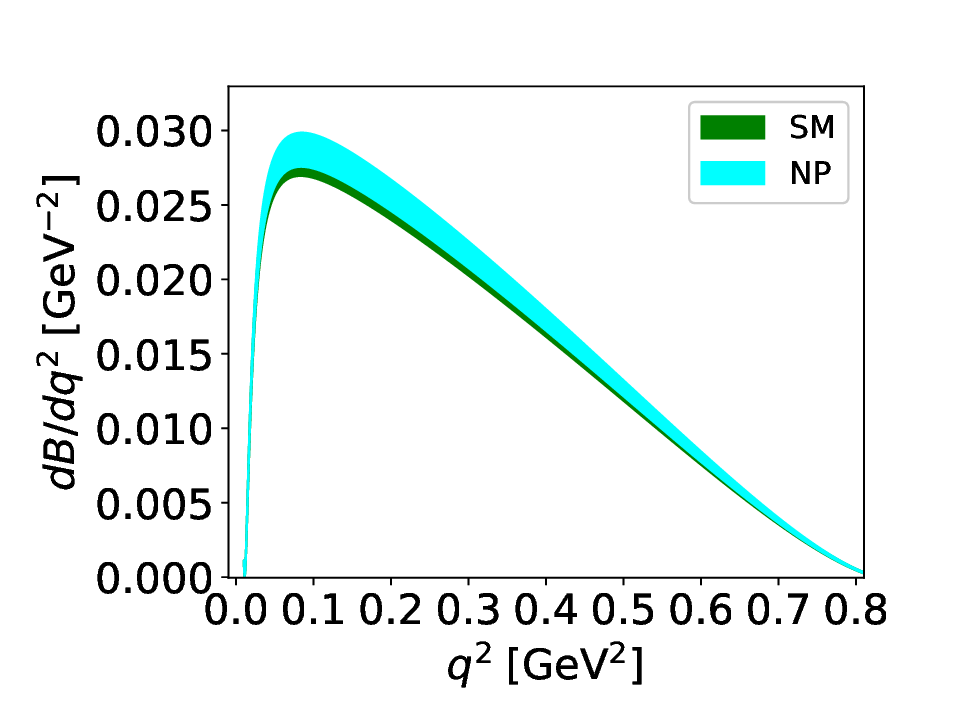}
	\includegraphics[width=0.22\textwidth]{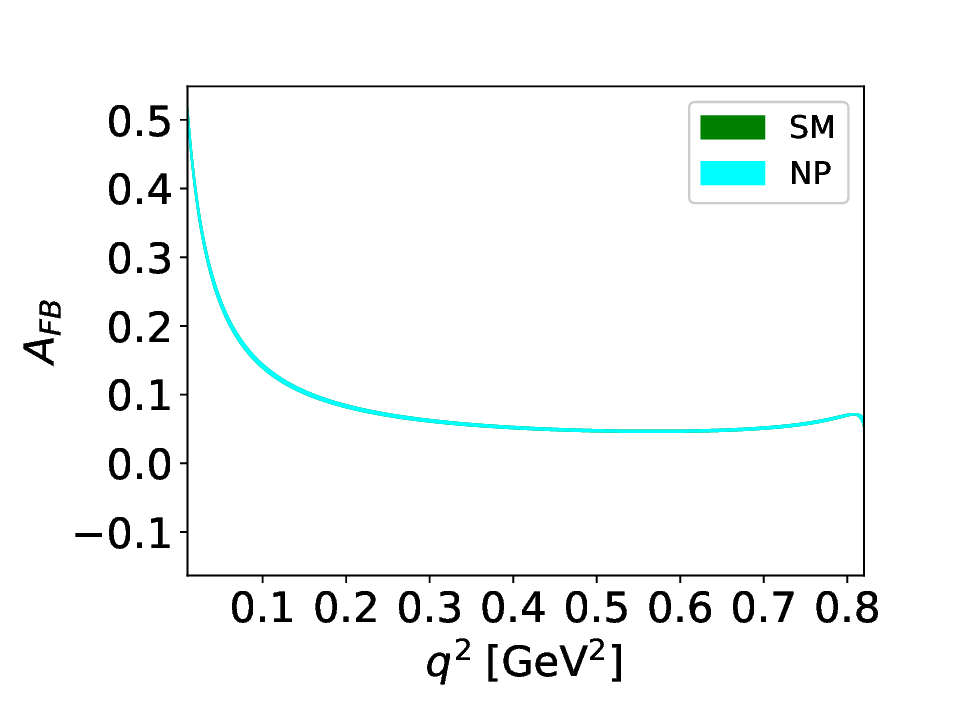}
	\includegraphics[width=0.22\textwidth]{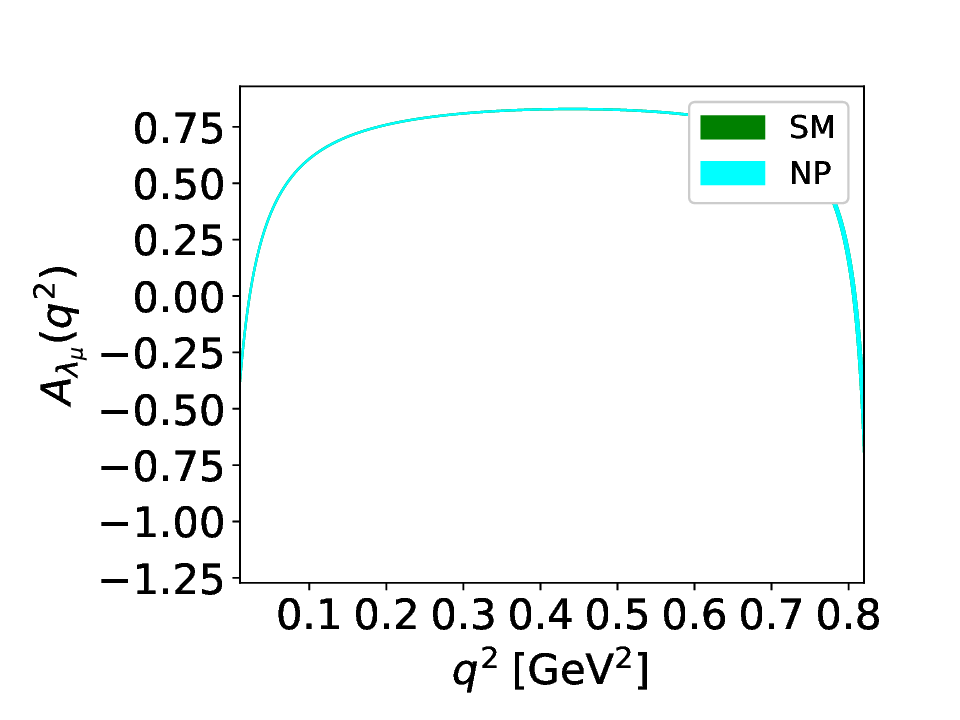}
	\includegraphics[width=0.22\textwidth]{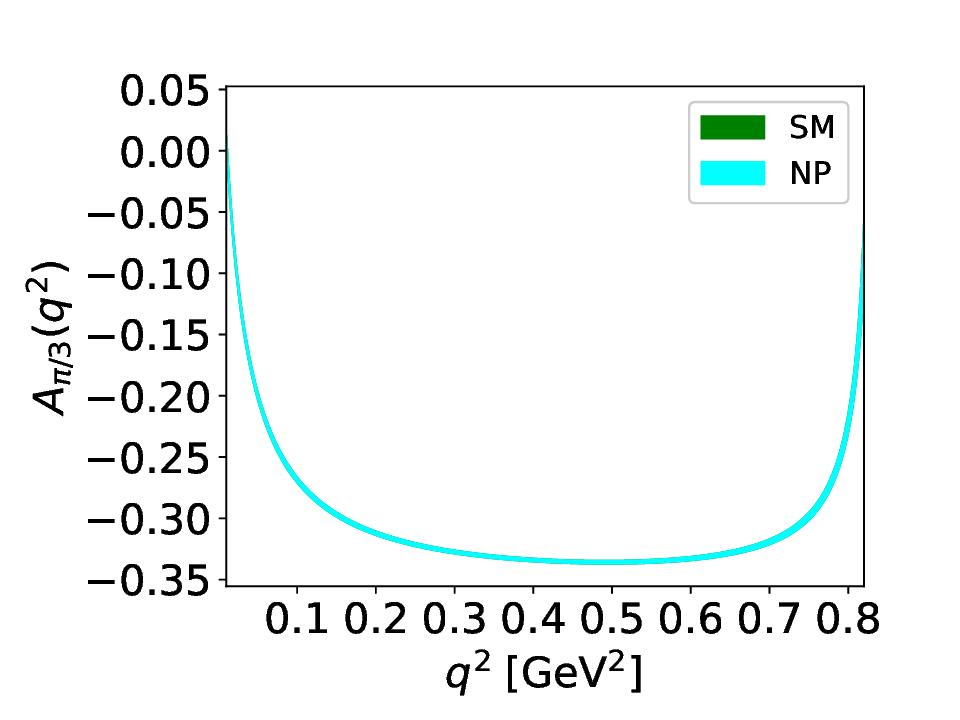}
	\caption{$q^2$ spectra of scenario 3 for the differential branching fraction $\frac{d\mathcal{B}}{dq^2}$ (top left), forward-backward asymmetry $A_{FB}$ (top right), lepton polarization asymmetry $A_{\lambda_{\mu}}$ (bottom left), and convexity $A_{\pi/3}$ (bottom right). The Standard Model (SM green band) and new physics (NP cyan band) both include uncertainty from the form factors.}
	\label{fig-5}       % Give a unique label
\end{figure}

The predictions for all the observables are shown in figure \ref{fig-5} and we can see that this scenario cannot provide any deviation from SM in any of the observables.

\subsection{Scenario 4a}
Scenario 4a is for the scalar boson mediator $\Phi$ with quantum numbers $(1,2,1/2)$ under $SU(3)_C \times SU(2)_L \times U(1)_Y$. This mediator generates the new physics scalar operators $\mathcal{O}_{LR}^S, \mathcal{O}_{RR}^S$. The fit values and respective correlation for this scenario are given as in Table \ref{tab-sc4a}.

\begin{table} [H]
	\caption{Scenario 4a: Fit values and correlation}
	\vspace{2pt}
	\setlength{\tabcolsep}{4pt}
	\label{tab-sc4a}
	\renewcommand{\arraystretch}{1.5}
	\centering\begin{tabular}{lc|ll}
		\hline
		\multirow{2}{*}{WCs}&\multirow{2}{*}{Fit Values}&\multicolumn{2}{c}{Correlation} \\        
		\cline{3-4}		
		&&$C_{LR}^S$&$C_{RR}^S$\\
		\hline	
		$C_{LR}^S$ &$-0.029 \pm 0.128$  & 1 & 0.98 \\		
		$C_{RR}^S$&$-0.035 \pm 0.128$ & 0.98 & 1 \\		
		\hline
	\end{tabular}
\end{table}

\begin{figure}%[H]
	\centering
	\includegraphics[width=0.22\textwidth]{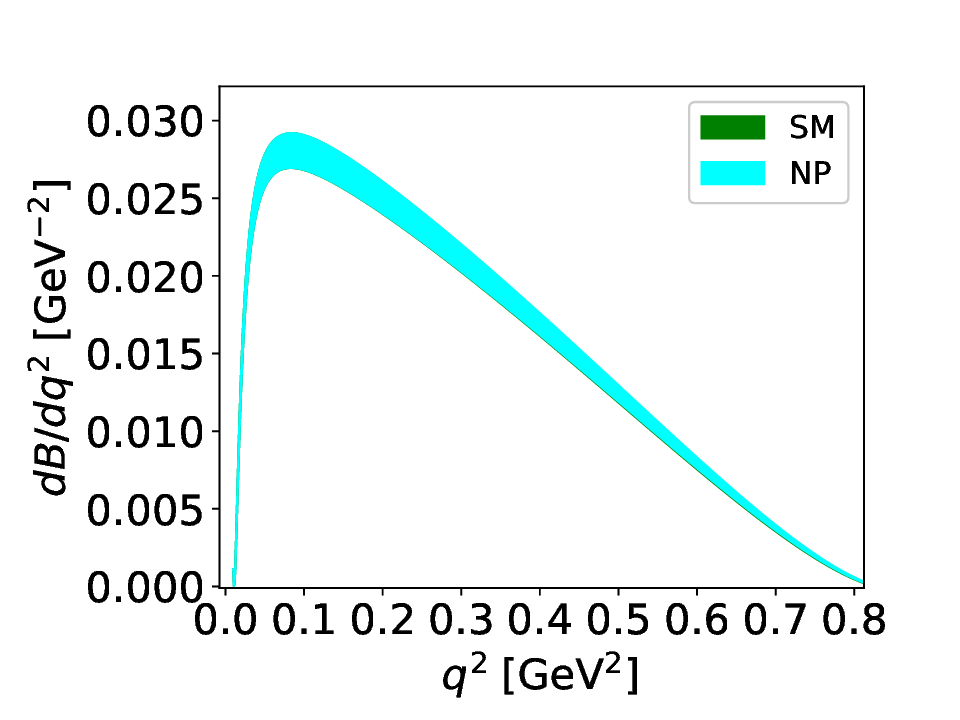}
	\includegraphics[width=0.22\textwidth]{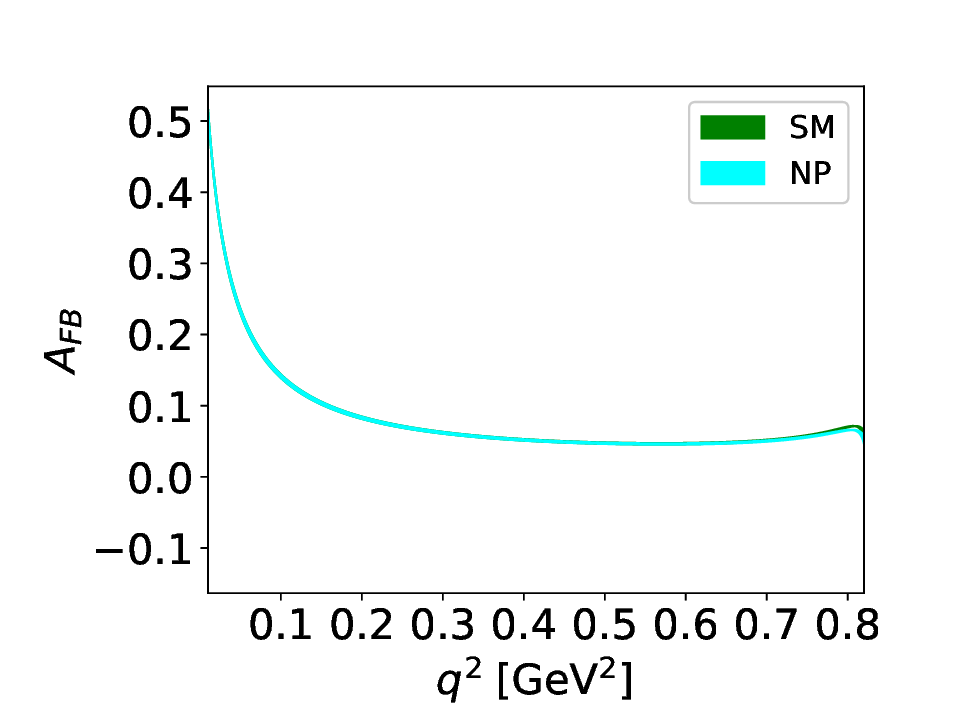}
	\includegraphics[width=0.22\textwidth]{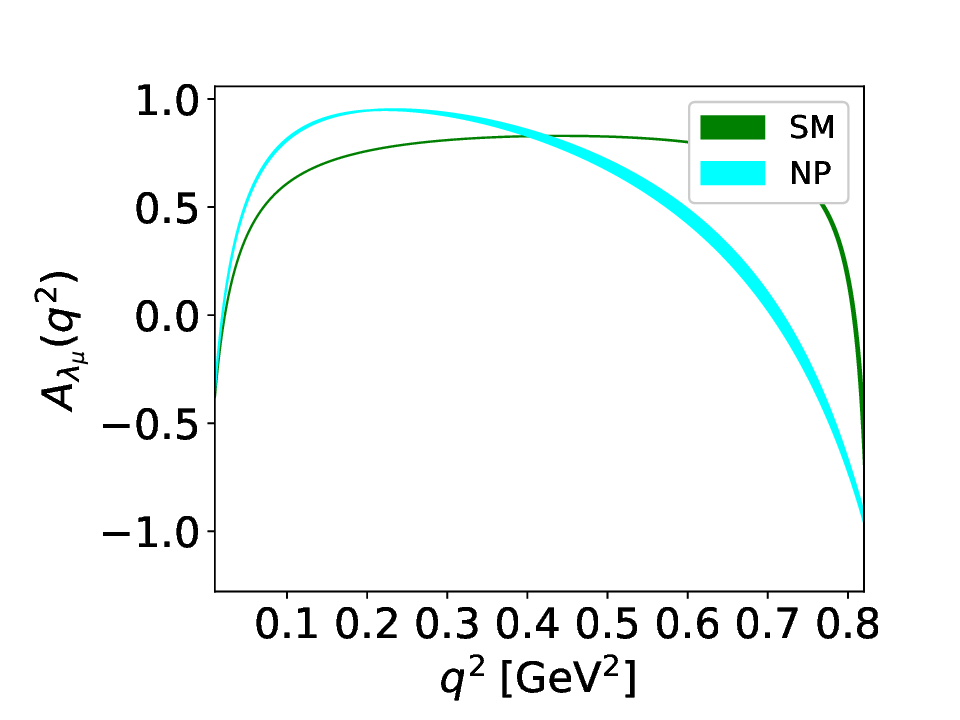}
	\includegraphics[width=0.22\textwidth]{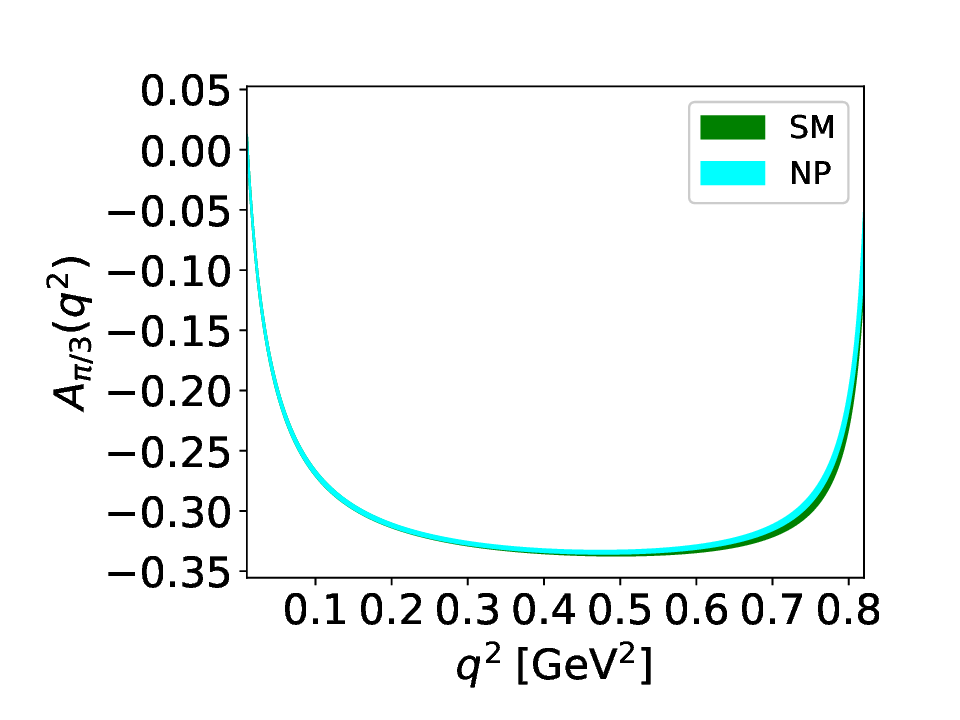}
	\caption{$q^2$ spectra of scenario 4a for the differential branching fraction $\frac{d\mathcal{B}}{dq^2}$ (top left), forward-backward asymmetry $A_{FB}$ (top right), lepton polarization asymmetry $A_{\lambda_{\mu}}$ (bottom left), and convexity $A_{\pi/3}$ (bottom right). The Standard Model (SM green band) and new physics (NP cyan band) both include uncertainty from the form factors.}
	\label{fig-6}       % Give a unique label
\end{figure}

As shown in figure \ref{fig-6}, lepton polarization asymmetry provides the deviation from the SM for the full range of $q^2$. The zero crossing for lepton polarization asymmetry shifts to a higher value of $q^2 \sim 0.7$ $GeV^2$. In the case of the other three observables, the $q^2$ spectrum is almost the same as SM.
\subsection{Scenario 4b}
This scenario is generated by including the operators having interactions with LHN in addition to RHN. This type of scenario can be generated from two Higgs doublet models where the second scalar doublet can have the interaction with LHN in addition to the RHN. The fit values and correlation for this scenario are listed as in Table \ref{tab-sc4b}.

\begin{table} [H]
	\caption{Scenario 4b: Fit values and correlation}
	\vspace{2pt}
	\setlength{\tabcolsep}{4pt}
	\label{tab-sc4b}
	\renewcommand{\arraystretch}{1.5}
	\centering\begin{tabular}{lc|llll}
		\hline
		\multirow{2}{*}{WCs}&\multirow{2}{*}{Fit Values}&\multicolumn{4}{c}{Correlation} \\        
		\cline{3-6}		
		&&$C_{LL}^S$ & $C_{RL}^S$ & $C_{LR}^S$&$C_{RR}^S$\\
		\hline	
		$C_{LL}^S$ &$-0.100 \pm 0.007$& 1 & 0.6 & 0.2 & -0.3 \\
		
		$C_{RL}^S$ &$-0.099 \pm 0.007$ &0.6 & 1 & -0.3 & 0.2\\
		
		$C_{LR}^S$ &$-0.075 \pm 0.052$ &0.2 & -0.3 & 1 & 0.3  \\
		
		$C_{RR}^S$&$-0.079 \pm 0.051$ & -0.3 &0.2 & 0.3 & 1 \\		
		\hline
	\end{tabular}
\end{table}

\begin{figure}[H]
	\centering
	\includegraphics[width=0.22\textwidth]{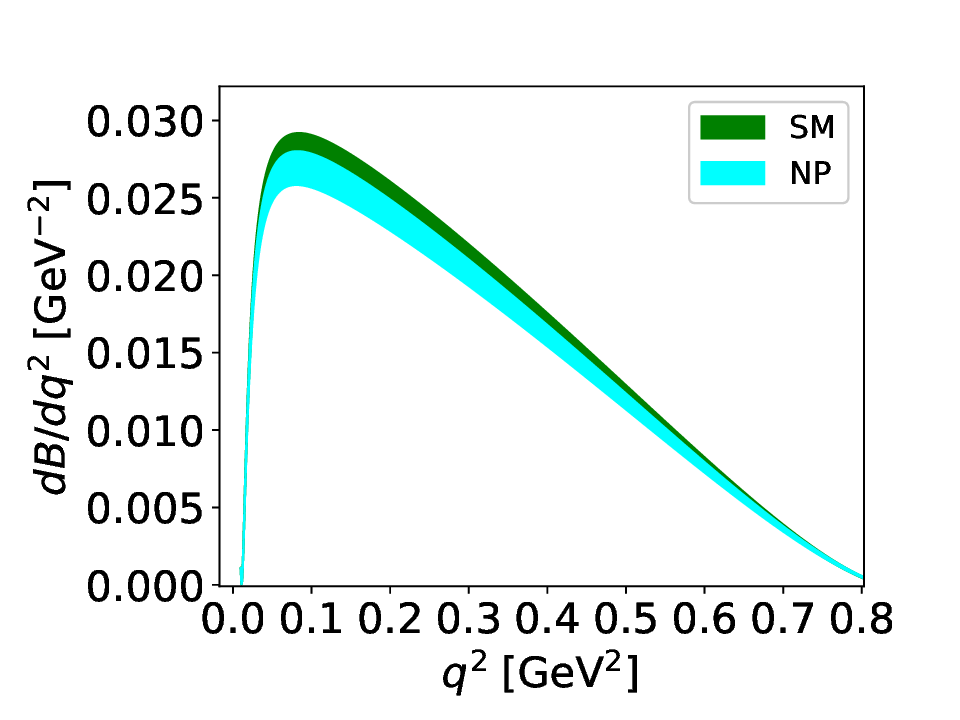}
	\includegraphics[width=0.22\textwidth]{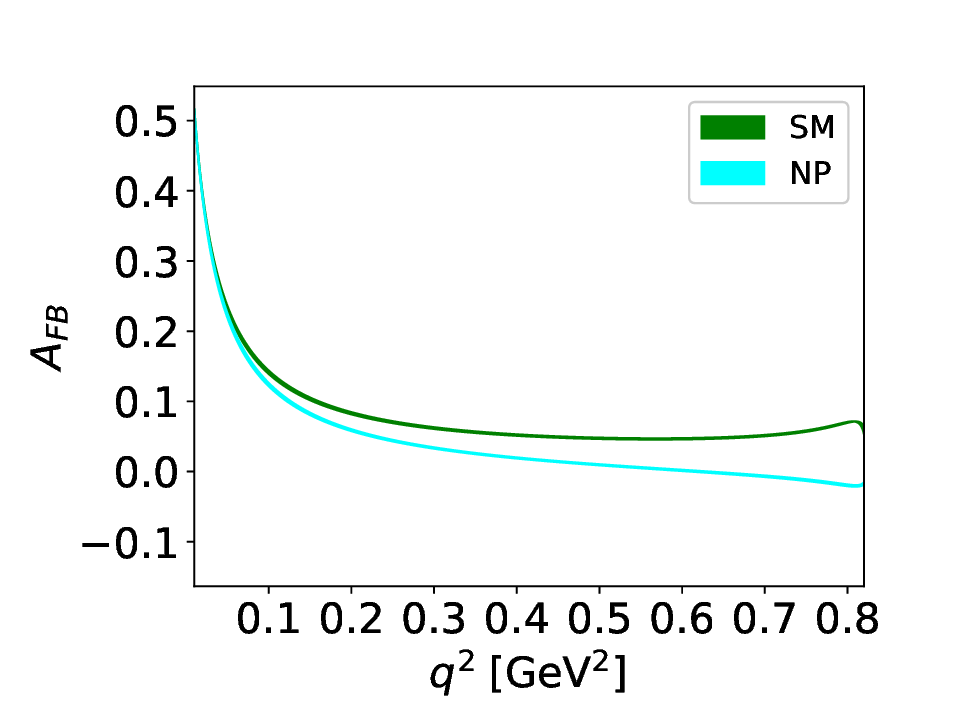}
	\includegraphics[width=0.22\textwidth]{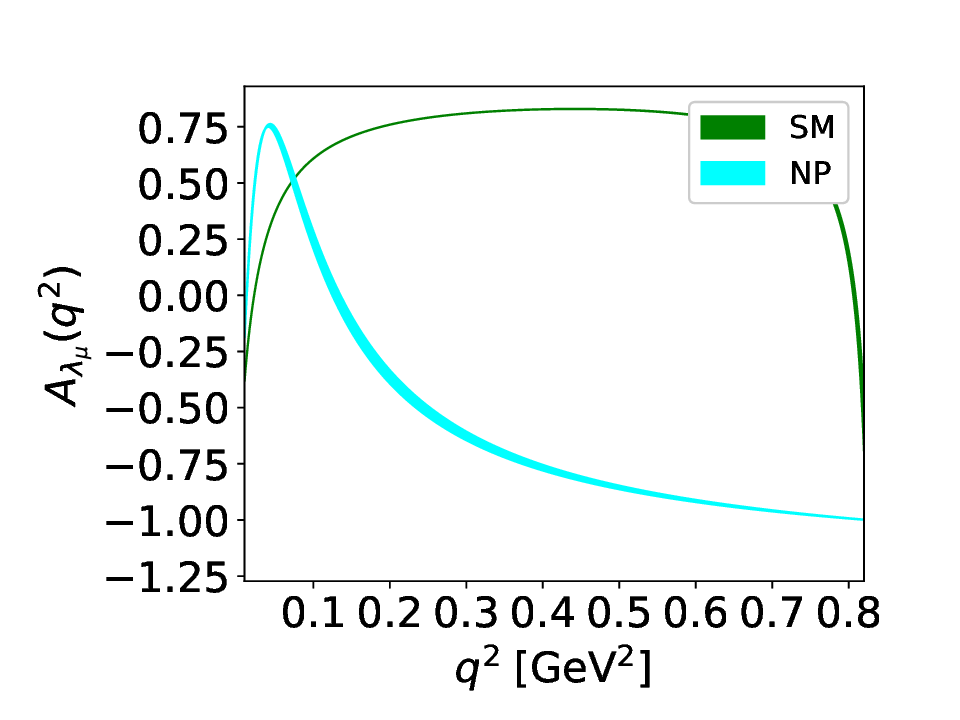}
	\includegraphics[width=0.22\textwidth]{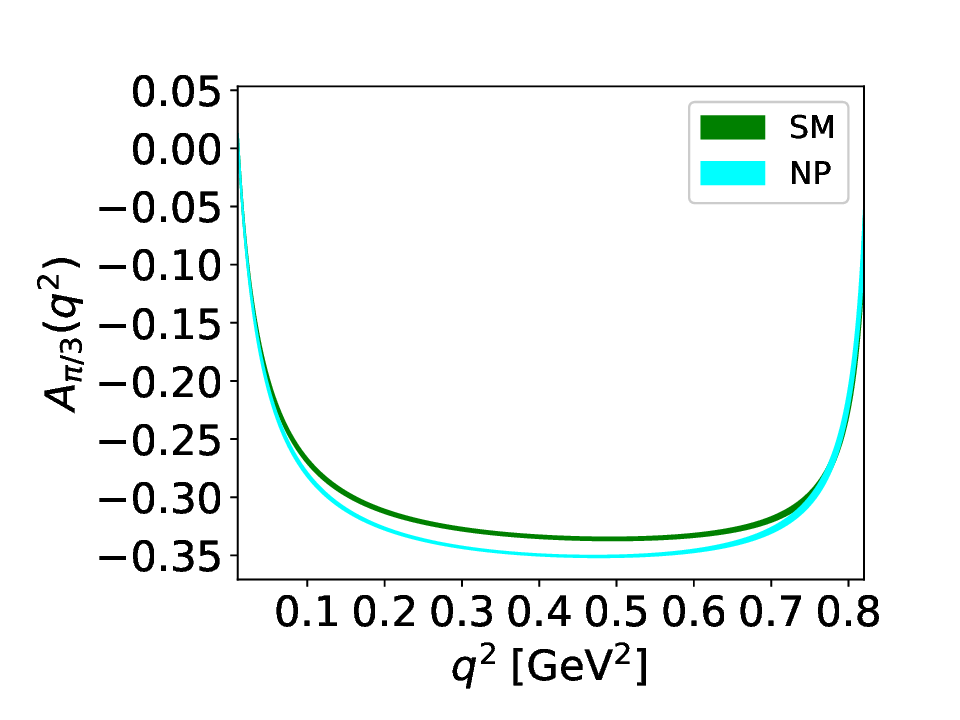}
	\caption{$q^2$ spectra of scenario 4b for the differential branching fraction $\frac{d\mathcal{B}}{dq^2}$ (top left), forward-backward asymmetry $A_{FB}$ (top right), lepton polarization asymmetry $A_{\lambda_{\mu}}$ (bottom left), and convexity $A_{\pi/3}$ (bottom right). The Standard Model (SM green band) and new physics (NP cyan band) both include uncertainty from the form factors.}
	\label{fig-7}       % Give a unique label
\end{figure}

As shown in figure \ref{fig-7}, in scenario 4b the convexity shows sensitivity for new physics for a $q^2$ range of (0.1 - 0.7)  $GeV^2$. Lepton polarization asymmetry shows a large deviation from the SM band and the zero crossing for NP is at lower value of $q^2 \sim 0.13 $ $ GeV^2$. The NP prediction of forward-backward aysmmetry also provide the deviation from SM as shown in the upper right panel of the figure. But the $q^2$ spectrum of differential branching fraction doesn't show any deviation from the SM in this scenario.

\subsection{Scenario 5a}
The vector leptoquark $U_1^{\mu}$ at a high energy scale can generate the left and right-handed operators at the $m_b$ scale. Considering the interaction with only RHNs,  the operators $\mathcal{O}_{RR}^V$, $\mathcal{O}_{LR}^S$ will contribute to the $B_c^+ \to B_s \mu^+ \nu_{\mu}$ process. We get the following fit values and correlation listed in Table \ref{tab-sc5a}

\begin{table} [H]
	\caption{Scenario 5a: Fit values and correlation}
	\vspace{2pt}
	\setlength{\tabcolsep}{5pt}
	\label{tab-sc5a}
	\renewcommand{\arraystretch}{1.5}
	\centering\begin{tabular}{lc|ll}
		\hline
		\multirow{2}{*}{WCs}&\multirow{2}{*}{Fit Values}&\multicolumn{2}{c}{Correlation} \\        
		\cline{3-4}		
		&&$C_{RR}^V$&$C_{LR}^S$\\
		\hline	
		$C_{RR}^V$ &$ 0.168 \pm 0.123$  & 1 & -0.8 \\		
		$C_{LR}^S$&$-0.012 \pm 0.005$ & -0.8 & 1 \\		
		\hline
	\end{tabular}
\end{table}

\begin{figure}[H]
	\centering
	\includegraphics[width=0.22\textwidth]{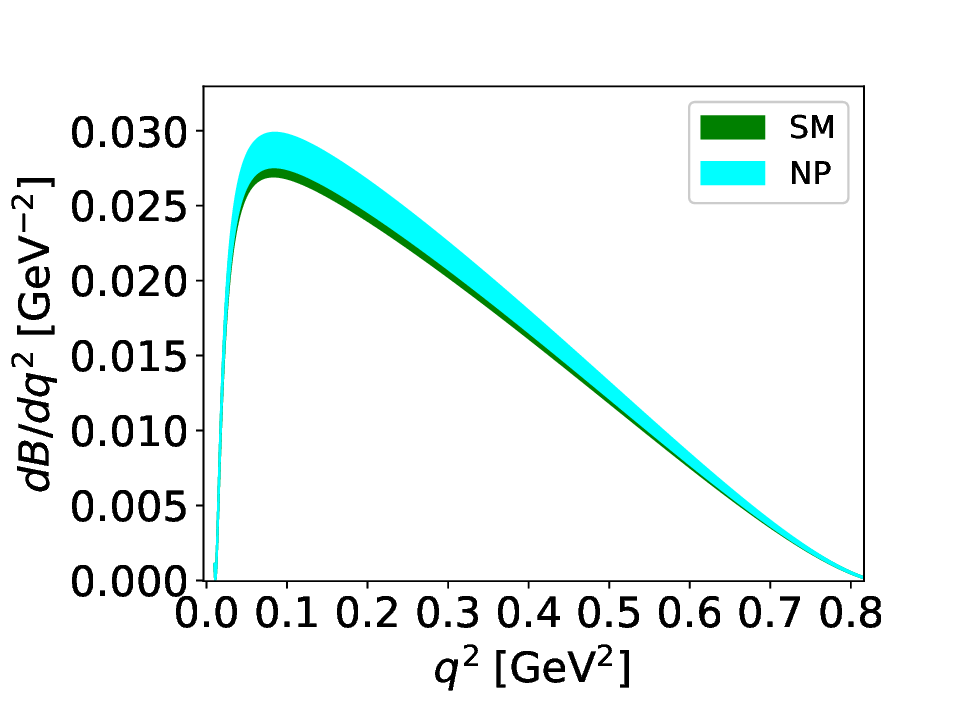}
	\includegraphics[width=0.22\textwidth]{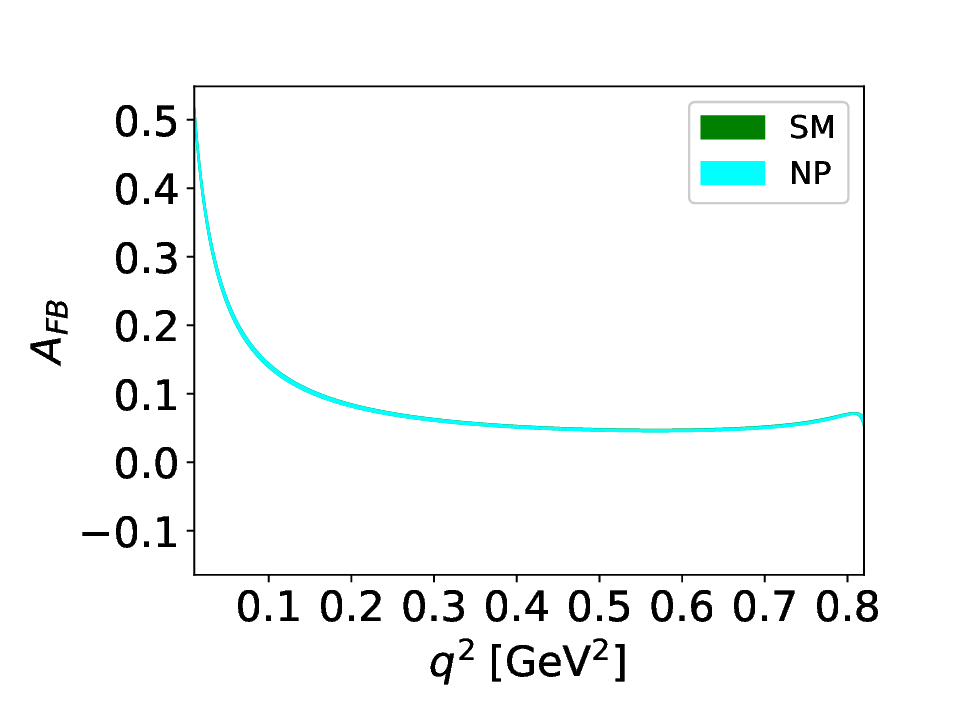}
	\includegraphics[width=0.22\textwidth]{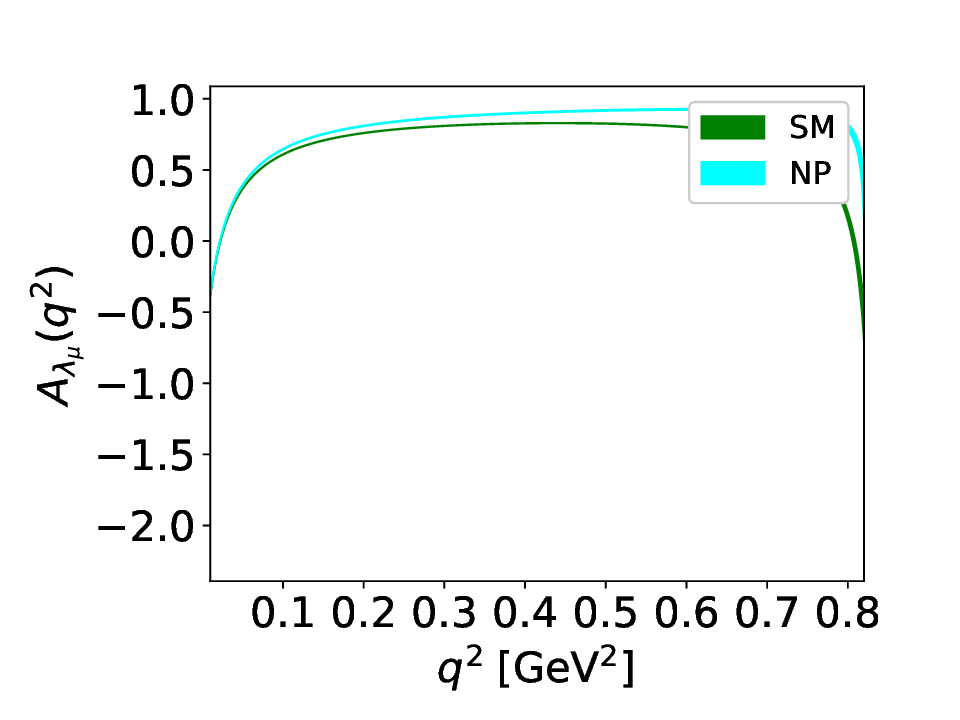}
	\includegraphics[width=0.22\textwidth]{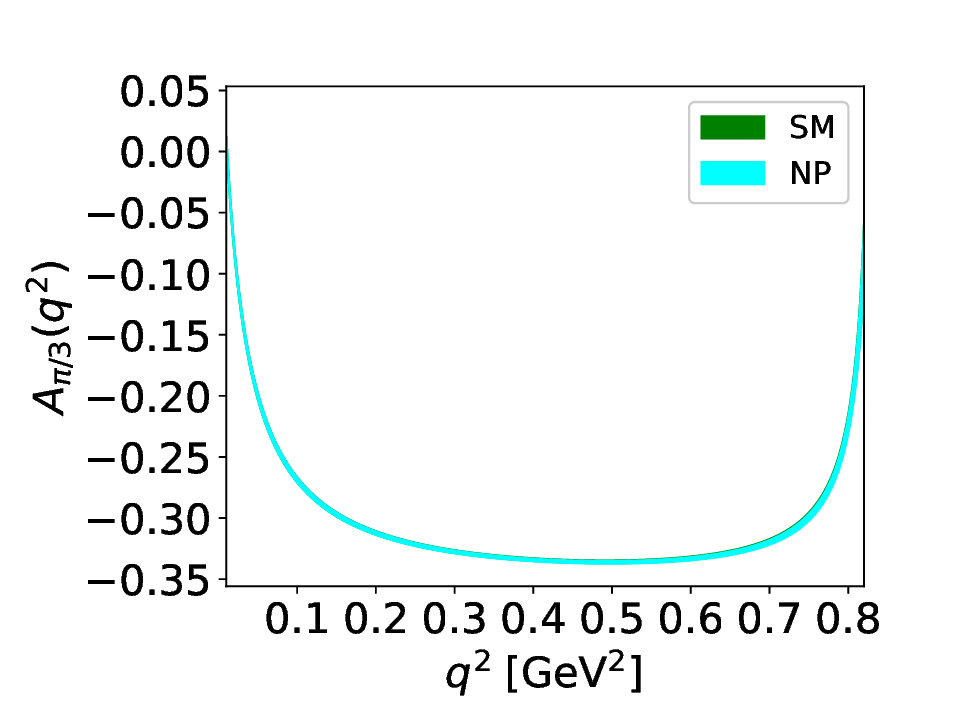}
	\caption{$q^2$ spectra of scenario 5a for the differential branching fraction $\frac{d\mathcal{B}}{dq^2}$ (top left), forward-backward asymmetry $A_{FB}$ (top right), lepton polarization asymmetry $A_{\lambda_{\mu}}$ (bottom left), and convexity $A_{\pi/3}$ (bottom right). The Standard Model (SM green band) and new physics (NP cyan band) both include uncertainty from the form factors.}
	\label{fig-8}       % Give a unique label
\end{figure}

As shown in figure \ref{fig-8}, only lepton polarization asymmetry can show the deviation from the SM. The other observables cannot provide any significant deviation from SM.

\subsection{Scenario 5b}
Including the interaction with LHN, this leptoquark provides the additional new physics operators $\mathcal{O}_{LL}^V$, $\mathcal{O}_{RL}^S$ in addition to the operators $\mathcal{O}_{RR}^V$, $\mathcal{O}_{LR}^S$ considered in scenarios 5a. The fit values and correlation between the operators in this scenario are as follows in Table \ref{tab-sc5b}.

\begin{table} [H]
	\caption{Scenario 5b: Fit values and correlation}
	\vspace{2pt}
	\setlength{\tabcolsep}{4pt}
	\label{tab-sc5b}
	\renewcommand{\arraystretch}{1.5}
	\centering\begin{tabular}{lc|llll}
		\hline
		\multirow{2}{*}{WCs}&\multirow{2}{*}{Fit Values}&\multicolumn{4}{c}{Correlation} \\        
		\cline{3-6}		
		&&$C_{LL}^V$ & $C_{RL}^S$ & $C_{LR}^S$&$C_{RR}^V$\\
		\hline	
		$C_{LL}^V$ &$-0.017 \pm 0.891$& 1 & -0.3 & 0.3 & -0.9 \\
		
	$C_{RL}^S$ &$-0.069 \pm 0.023$ &-0.3 & 1 & -0.9 & 0.3\\
		
		$C_{LR}^S$ &$-0.019 \pm 0.037$ &0.3 & -0.9 & 1 & -0.3  \\
		
		$C_{RR}^V$&$ 0.283 \pm 3.088$ & -0.9 & 0.3 & -0.3 & 1 \\		
		\hline
	\end{tabular}
\end{table}

\begin{figure}%[H]
	\centering
	\includegraphics[width=0.22\textwidth]{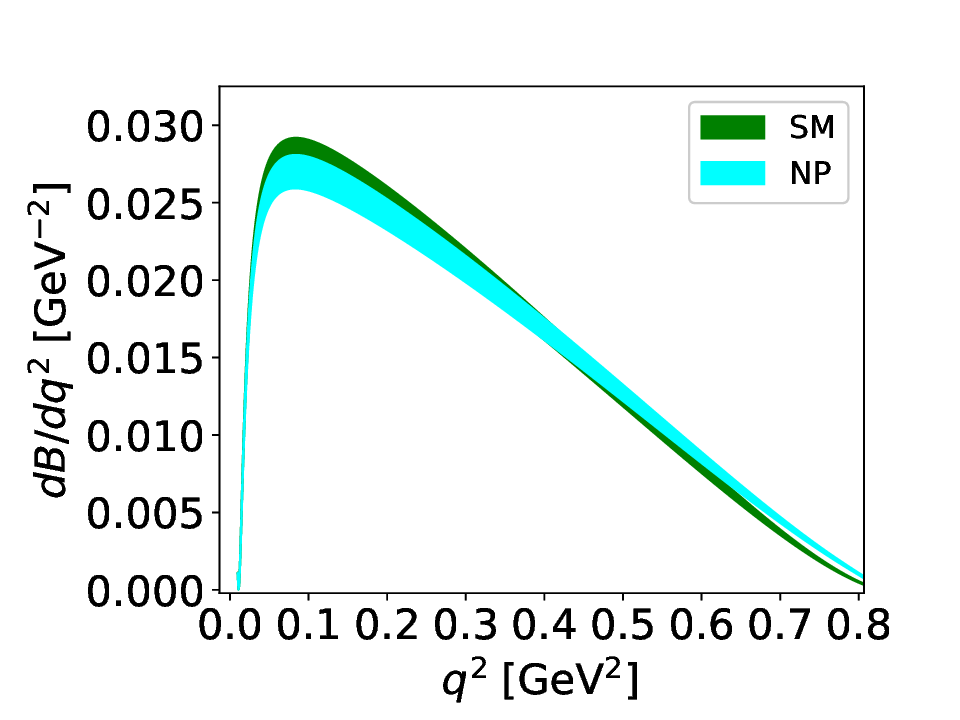}
	\includegraphics[width=0.22\textwidth]{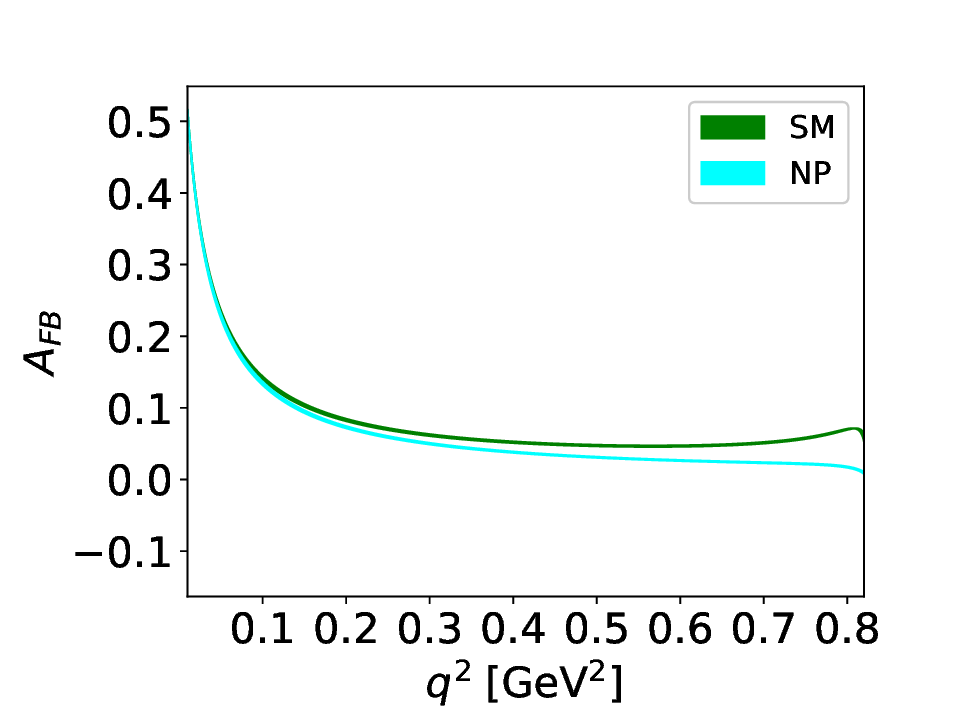}
	\includegraphics[width=0.22\textwidth]{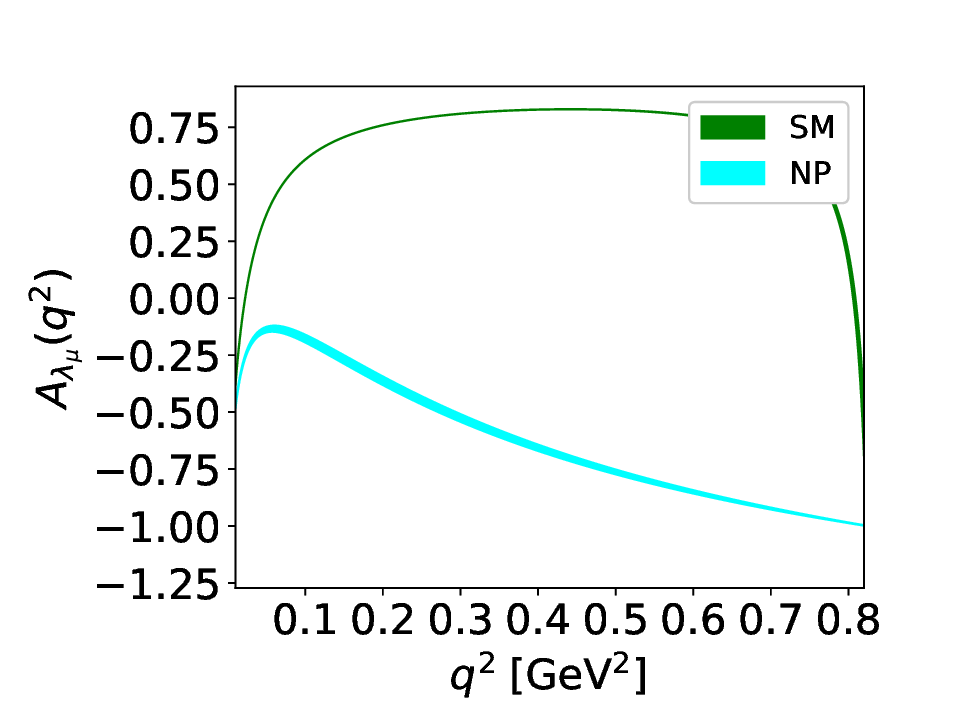}
	\includegraphics[width=0.22\textwidth]{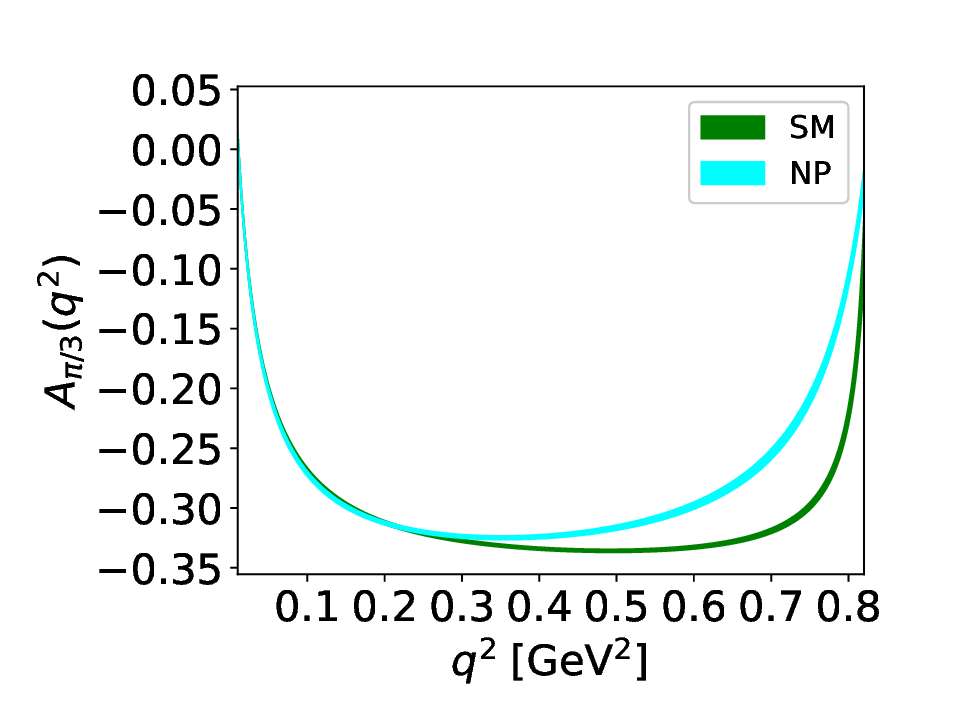}
	\caption{$q^2$ spectra of scenario 5b for the differential branching fraction $\frac{d\mathcal{B}}{dq^2}$ (top left), forward-backward asymmetry $A_{FB}$ (top right), lepton polarization asymmetry $A_{\lambda_{\mu}}$ (bottom left), and convexity $A_{\pi/3}$ (bottom right). The Standard Model (SM green band) and new physics (NP cyan band) both includes uncertainty from the form factors.}
	\label{fig-9}       % Give a unique label
\end{figure}

In figure \ref{fig-9}, we find that the forward-backward asymmetry and convexity both provide a significant deviation from the SM for $q^2 > 0.3$ $GeV^2$ values. In the case of the lepton polarization asymmetry, we observed one interesting point that NP band lies in the negative axis for the whole $q^2$ range. The differential branching fraction parameter does not provide any deviation from the SM prediction.

\subsection{Sceanrio 6}
The scalar leptoquark $\tilde{R}_2 (3,2,1/6)$ provides this scenario with interaction only to RHN. This mediator does not interact with the LHN. The contribution in this scenario arises from the tensor operator $\mathcal{O}_{RR}^T$ and scalar operator $\mathcal{O}_{RR}^S$ which are related through Fierz identity $C_{RR}^S = 4rC_{RR}^T$ \footnote{where $r \sim 2$ at b-quark scale \cite{Mandal:2020htr}}. The fit value for this scenario are as follows:

	\begin{equation}
	\label{equ-sc6fv}
	C_{RR}^T = -0.070 \pm 0.032
	\end{equation}

\begin{figure}%[H]
	\centering
	\includegraphics[width=0.22\textwidth]{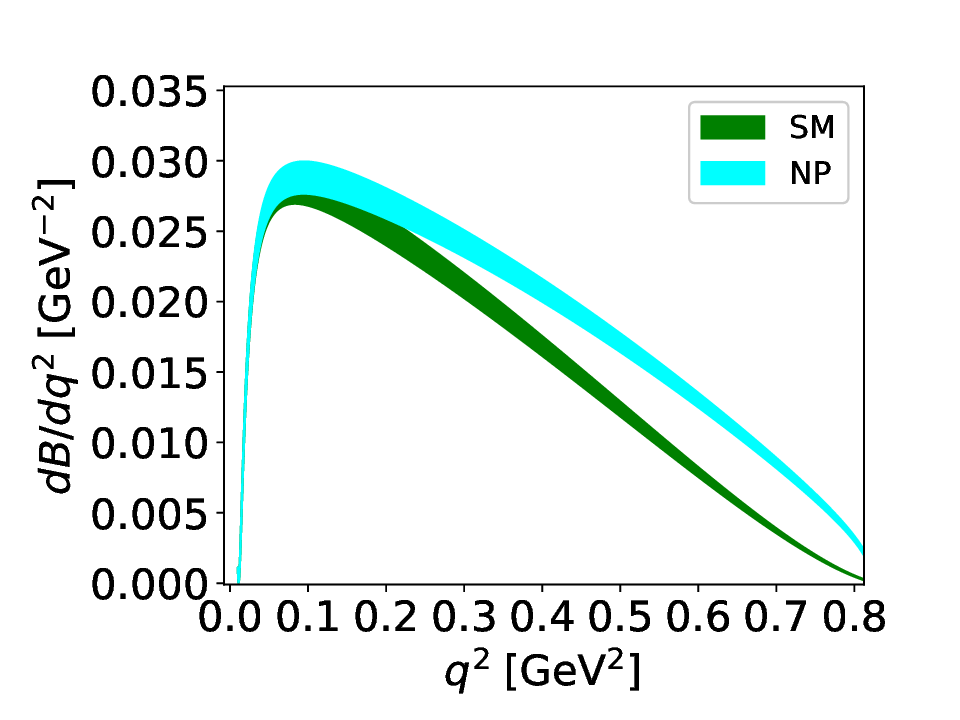}
	\includegraphics[width=0.22\textwidth]{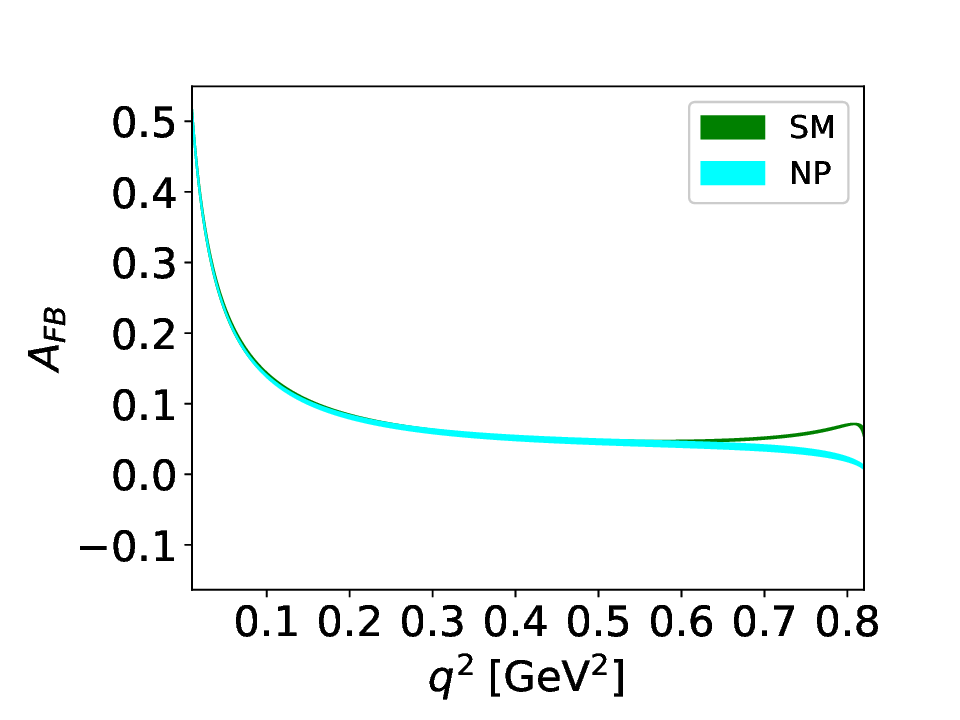}
	\includegraphics[width=0.22\textwidth]{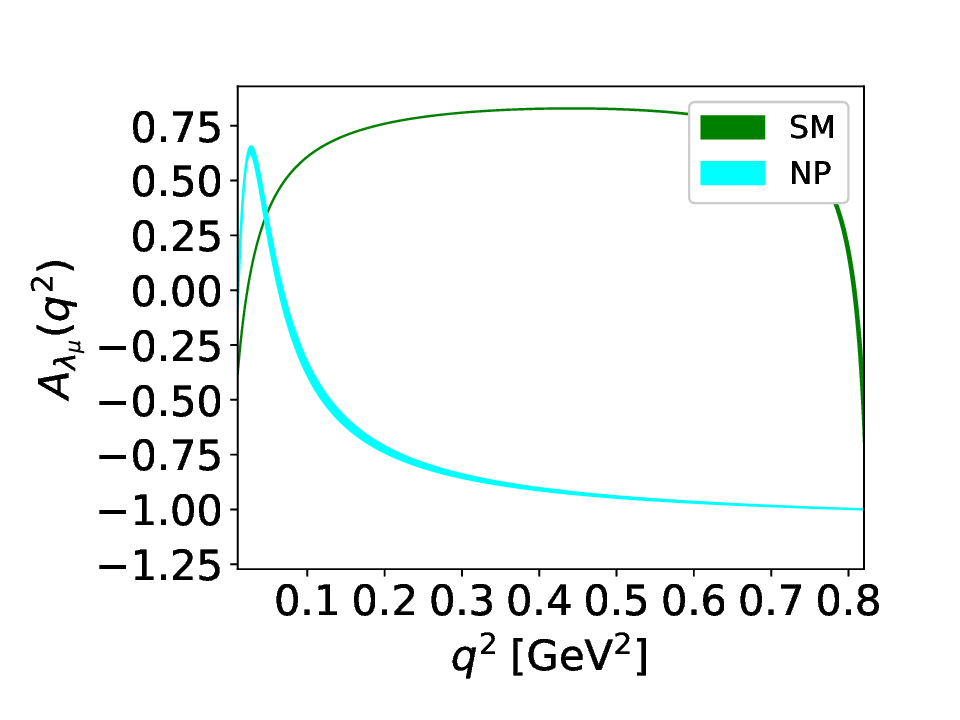}
	\includegraphics[width=0.22\textwidth]{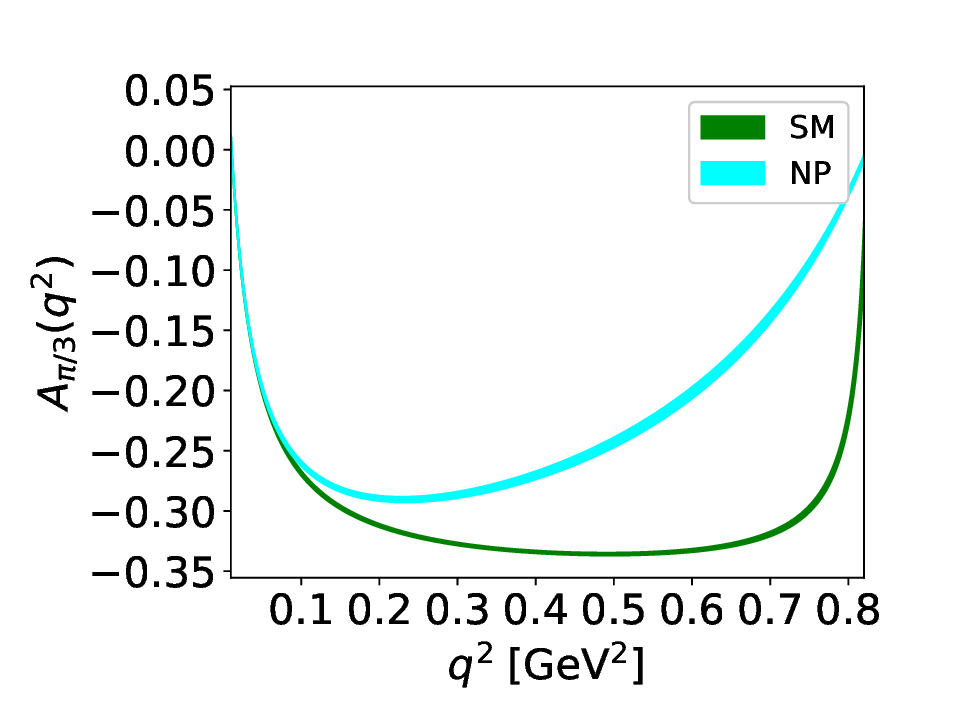}
	\caption{$q^2$ spectra of scenario 6 for the differential branching fraction $\frac{d\mathcal{B}}{dq^2}$ (top left), forward-backward asymmetry $A_{FB}$ (top right), lepton polarization asymmetry $A_{\lambda_{\mu}}$ (bottom left), and convexity $A_{\pi/3}$ (bottom right). The Standard Model (SM green band) and new physics (NP cyan band) both include uncertainty from the form factors.}
	\label{fig-10}       % Give a unique label
\end{figure}

For this scenario as shown in figure \ref{fig-10}, the differential branching fraction provides a deviation from SM above $q^2 \sim 0.2$ $GeV^2$. The lepton polarization asymmetry and convexity parameters give a large deviation from the SM almost for the full $q^2$ region. The zero crossing for lepton polarization asymmetry also shifts towards lower value of $q^2 \sim 0.08$ $ GeV^2$. The predictions for forward-backward asymmetry are similar to SM except beyond $q^2 \sim 0.72$ $ GeV^2$.

\subsection{Scenario 7a}
We consider the scalar leptoquark $S_1(\bar{3},1,1/3)$ which generates the vector ($\mathcal{O}_{RR}^V$), scalar ($\mathcal{O}_{RR}^S$), and tensor ($\mathcal{O}_{RR}^T$) operators with RHN interactions. The operators $\mathcal{O}_{RR}^S$ and $\mathcal{O}_{RR}^T$ are related as $C_{RR}^S = - 4rC_{RR}^T$. The fit values of the operartors with correlation are listed as in Table \ref{tab-sc7a}.

\begin{table} [H]
	\caption{Scenario 7a: Fit values and correlation}
	\vspace{2pt}
	\setlength{\tabcolsep}{5pt}
	\label{tab-sc7a}
	\renewcommand{\arraystretch}{1.5}
	\centering\begin{tabular}{lc|ll}
		\hline
		\multirow{2}{*}{WCs}&\multirow{2}{*}{Fit Values}&\multicolumn{2}{c}{Correlation} \\        
		\cline{3-4}		
		&&$C_{RR}^V$&$C_{RR}^T$\\
		\hline	
		$C_{RR}^V$ &$-0.134 \pm 0.102$  &1 & -0.3  \\		
		$C_{RR}^T$&$ -0.067 \pm 0.039$ & -0.3 &1 \\		
		\hline
	\end{tabular}
\end{table}

\begin{figure}%[H]
	\centering
	\includegraphics[width=0.22\textwidth]{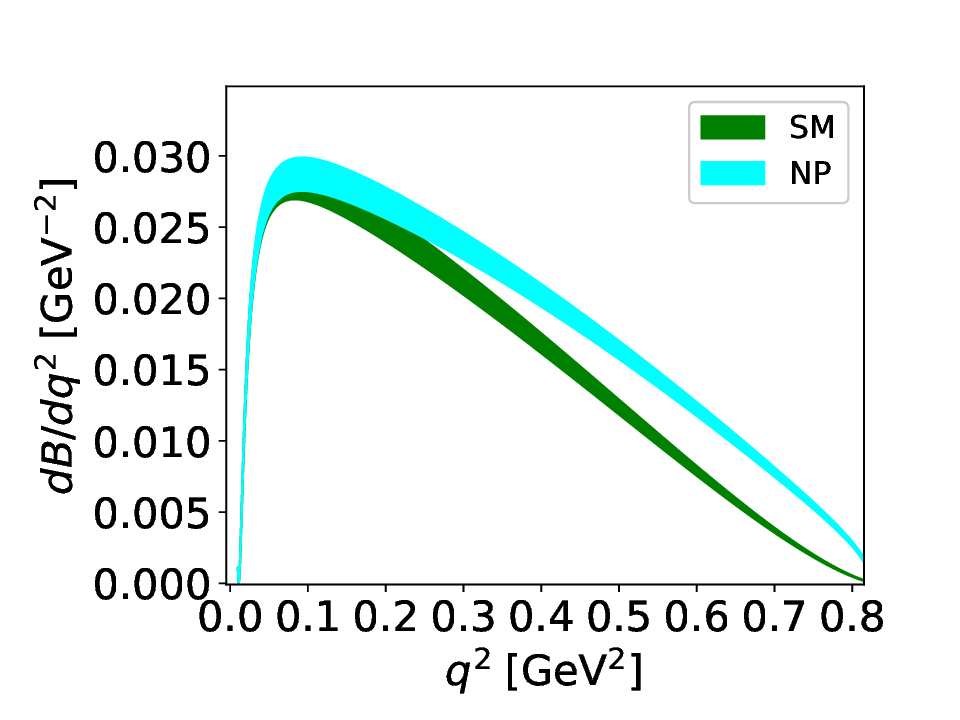}
	\includegraphics[width=0.22\textwidth]{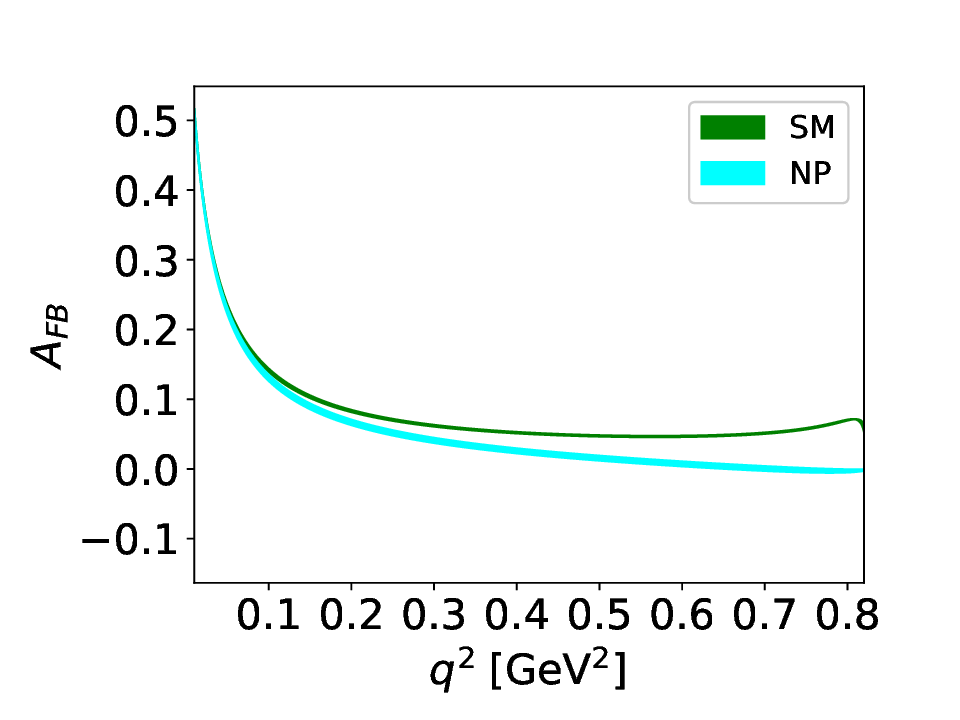}
	\includegraphics[width=0.22\textwidth]{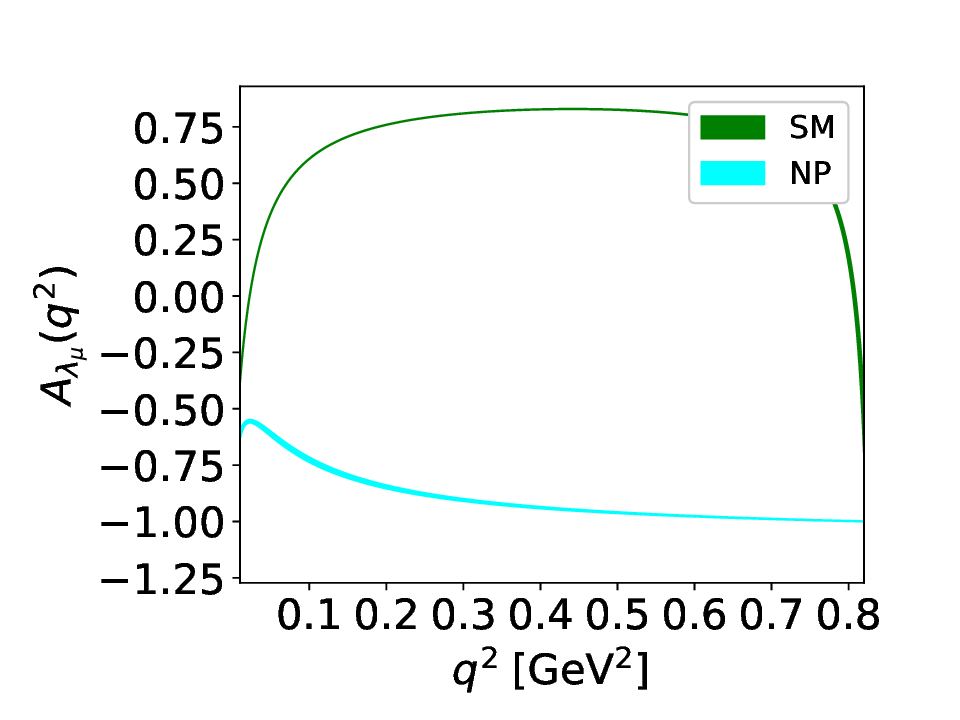}
	\includegraphics[width=0.22\textwidth]{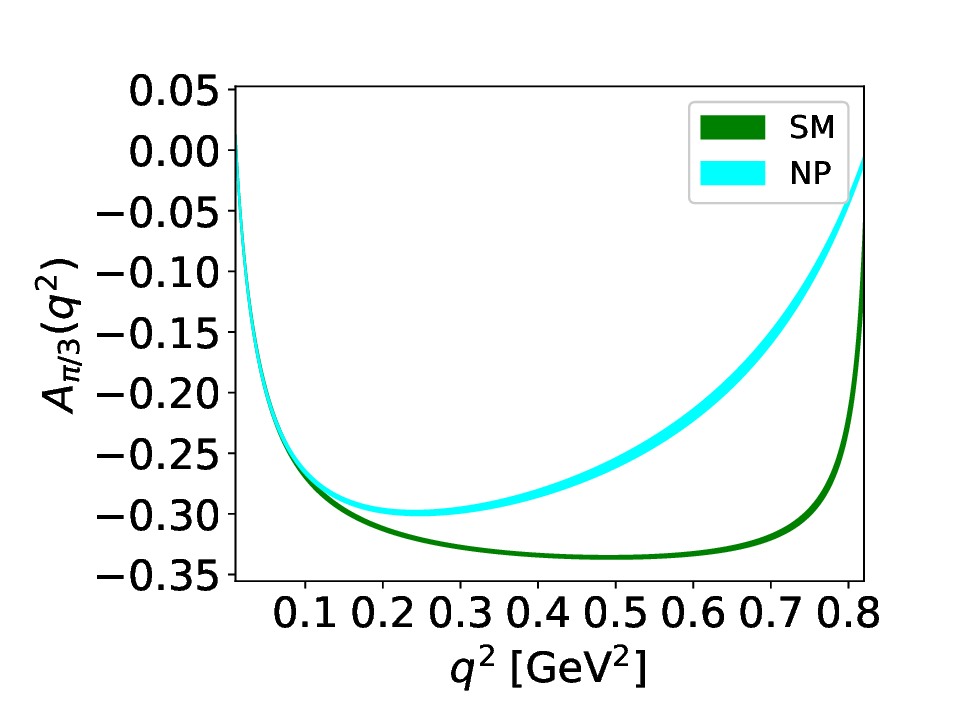}
	\caption{$q^2$ spectra of scenario 7a for the differential branching fraction $\frac{d\mathcal{B}}{dq^2}$ (top left), forward-backward asymmetry $A_{FB}$ (top right), lepton polarization asymmetry $A_{\lambda_{\mu}}$ (bottom left), and convexity $A_{\pi/3}$ (bottom right). The Standard Model (SM green band) and new physics (NP cyan band) both include uncertainty from the form factors.}
	\label{fig-11}       % Give a unique label
\end{figure}

As shown in figure \ref{fig-11}, all four observables show the sensitivity for this scenario. The differential branching fraction and forward-backward asymmetry provide a deviation from SM for $q^2 > 0.2$ $ GeV^2$. The lepton polarization asymmetry gives a large deviation from SM and provides the negative values of this observable in the full $q^2$ range. The convexity parameter also gives a large deviation from SM with a $q^2$ range of beyond the 0.18 $GeV^2$.

\subsection{Sceanrio 7b}
The scenario 7b arises with the same leptoquark considered in scenario 7a allowing the interaction with LHN to provide the additional new physics operators $\mathcal{O}_{LL}^V$, $\mathcal{O}_{LL}^S$ and $\mathcal{O}_{LL}^T$ in addition to the operators considered in scenario 7a. Also, the left-handed scalar operator is related to the tensor operator by Fierz identity $C_{LL}^S = - 4rC_{LL}^T$. The fit values and correlation for all the operators are listed in Table \ref{tab-sc7b}.

\begin{table}[H]
	\caption{Scenario 7b: Fit values and correlation}
	\vspace{2pt}
	\setlength{\tabcolsep}{3pt}
	\label{tab-sc7b}
	\renewcommand{\arraystretch}{1.5}
	\centering\begin{tabular}{lc|llll}
		\hline
		\multirow{2}{*}{WCs}&\multirow{2}{*}{Fit Values}&\multicolumn{4}{c}{Correlation} \\        
		\cline{3-6}		
		&&$C_{LL}^V$ & $C_{LL}^T$ & $C_{RR}^V$&$C_{RR}^T$\\
		\hline	
	$C_{LL}^V$ &$-0.092 \pm 0.208$& 1 & 0 & 0.95 & -0.1 \\
		
	$C_{LL}^T$ &$-0.026 \pm 0.092$ &0 & 1 & -0.1 & 0.9\\
		
		$C_{RR}^V$ &$-0.448 \pm 0.421$ &0.95 & -0.1 & 1 & -0.1  \\
		
	$C_{RR}^T$&$ 0.012 \pm 0.183$ & -0.1 & 0.9 & -0.1 & 1 \\		
		\hline
	\end{tabular}
\end{table}

\begin{figure}[H]
	\centering
	\includegraphics[width=0.22\textwidth]{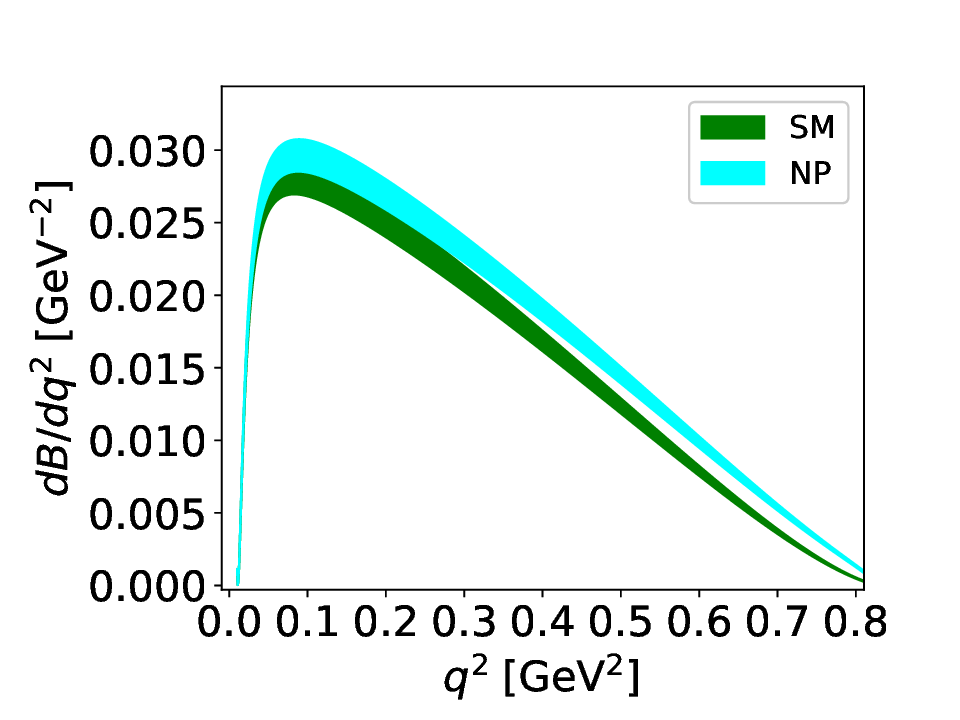}
	\includegraphics[width=0.22\textwidth]{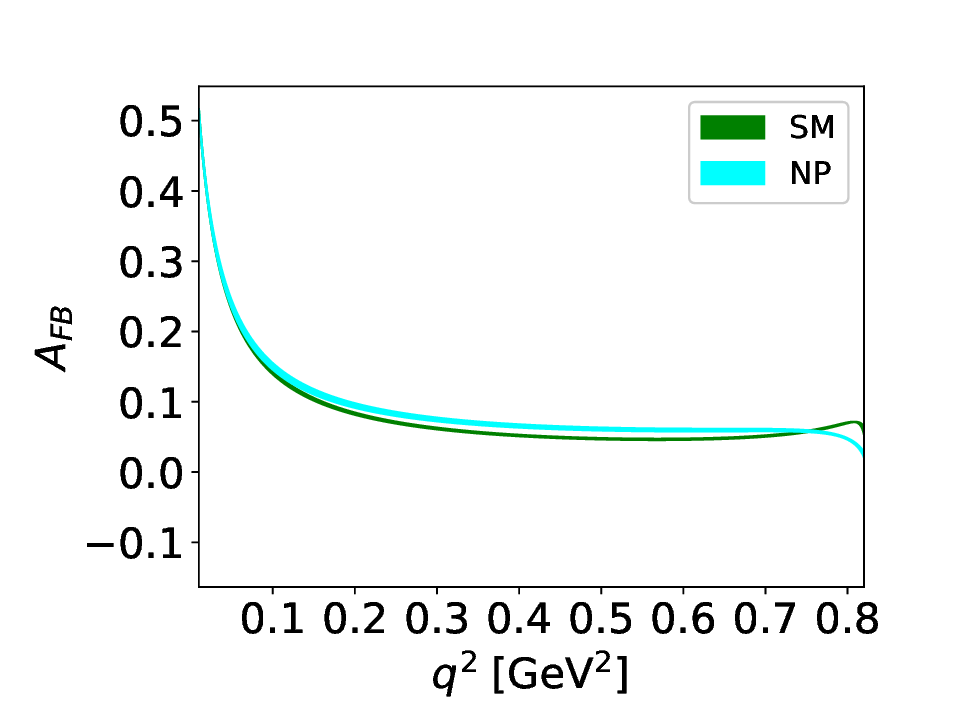}
	\includegraphics[width=0.22\textwidth]{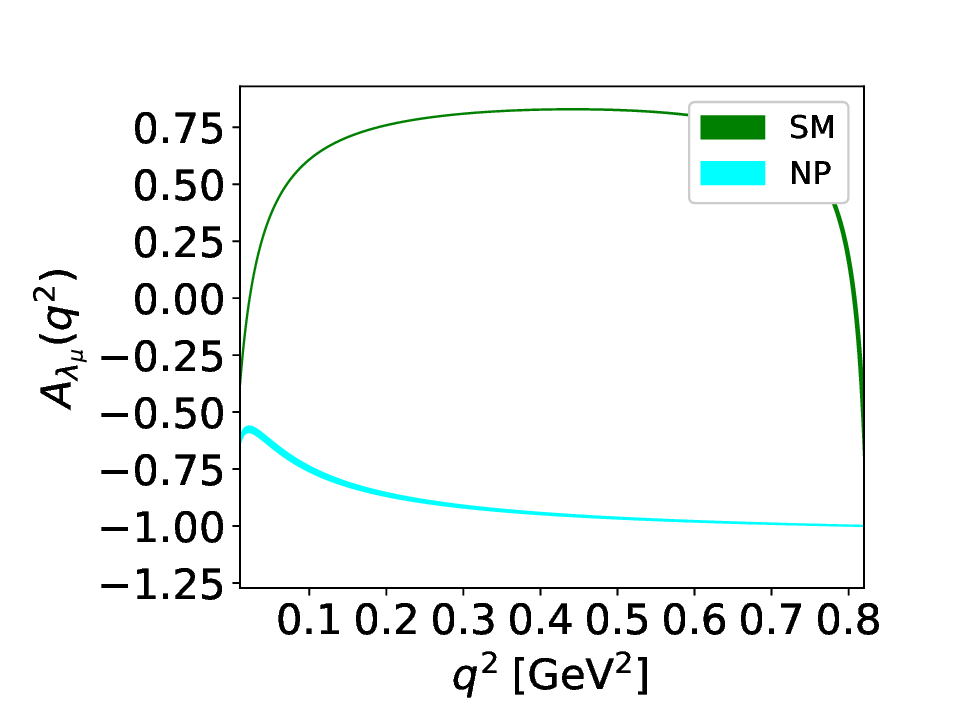}
	\includegraphics[width=0.22\textwidth]{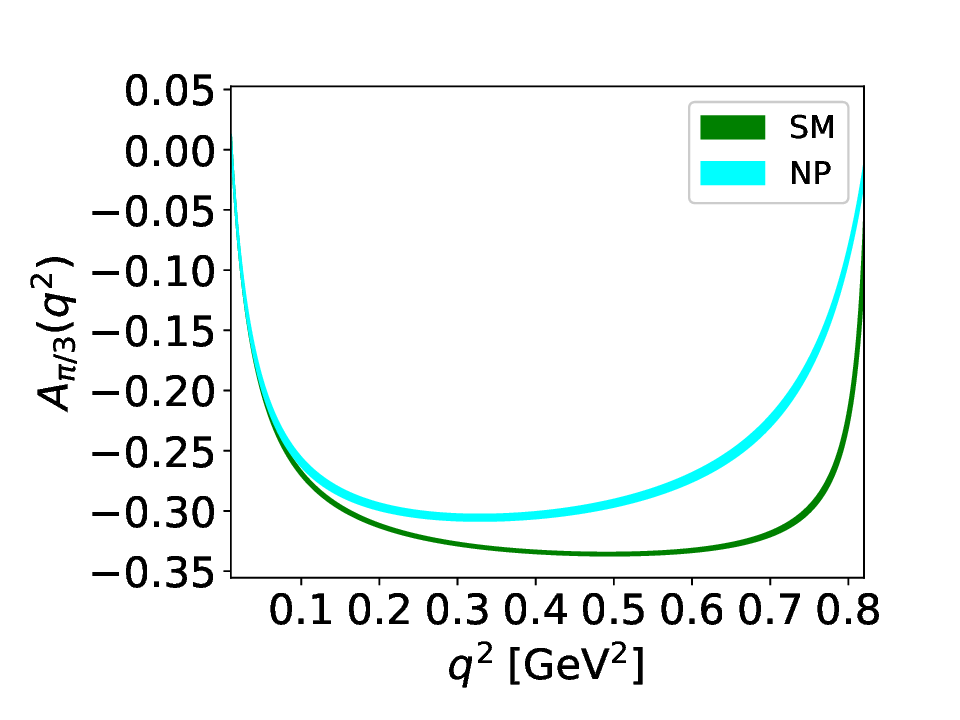}
	\caption{$q^2$ spectra of scenario 7b for the differential branching fraction $\frac{d\mathcal{B}}{dq^2}$ (top left), forward-backward asymmetry $A_{FB}$ (top right), lepton polarization asymmetry $A_{\lambda_{\mu}}$ (bottom left), and convexity $A_{\pi/3}$ (bottom right). The Standard Model (SM green band) and new physics (NP cyan band) both include uncertainty from the form factors.}
	\label{fig-12}       % Give a unique label
\end{figure}

The predictions for the $q^2$ distribution for the considered observables in this study are shown in figure \ref{fig-12} in scenario 7b. From the figure, we can see that the differential branching fraction and forward-backward asymmetry show a small deviation from SM. However, the lepton polarization asymmetry and convexity both are showing strong sensitivity for the NP operators. The lepton polarization asymmetry for NP scenario provides a $q^2$ spectrum that is negative in the full $q^2$ range.

\subsection{Sceanrio 8}
We consider the vector leptoquark $\tilde{V}_2^{\mu}$ with quantum numbers $(\bar{3}, 2, -1/6)$. This is a genuine scenario of the RHNs as this leptoquark does not have interaction with left-handed neutrinos. It provides the new physics contribution to $c \to s\ell \nu$ through the operator $\mathcal{O}_{LR}^S$. The fit value is given as

	\begin{equation}
	\label{equ-sc8fv}
	C_{LR}^S = -0.006 \pm 0.003
	\end{equation}

\begin{figure}%[H]
	\centering
	\includegraphics[width=0.22\textwidth]{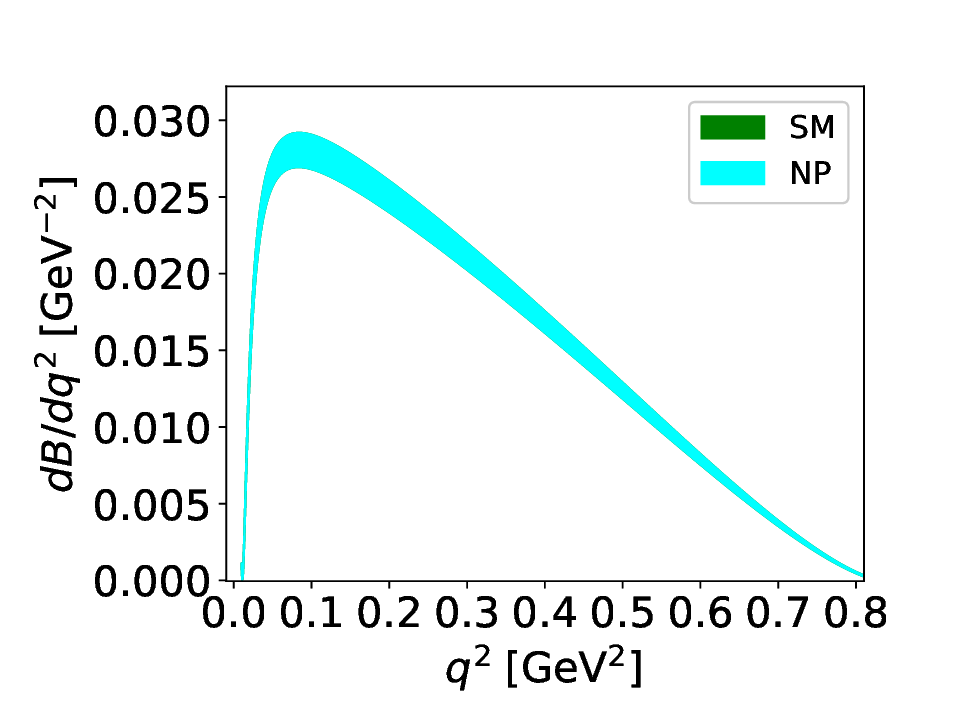}
	\includegraphics[width=0.22\textwidth]{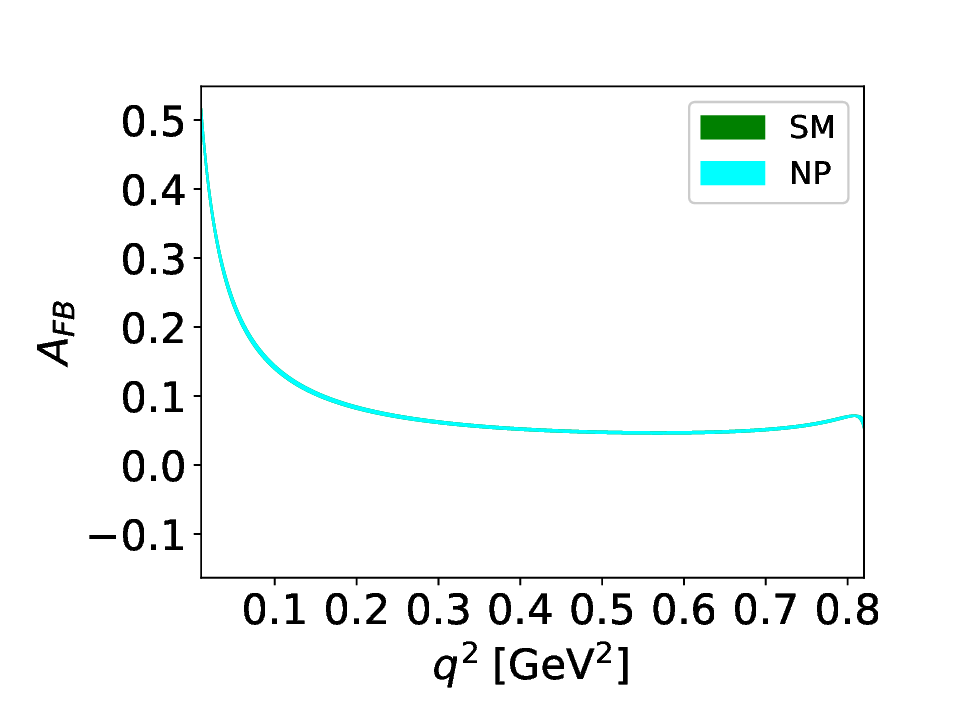}
	\includegraphics[width=0.22\textwidth]{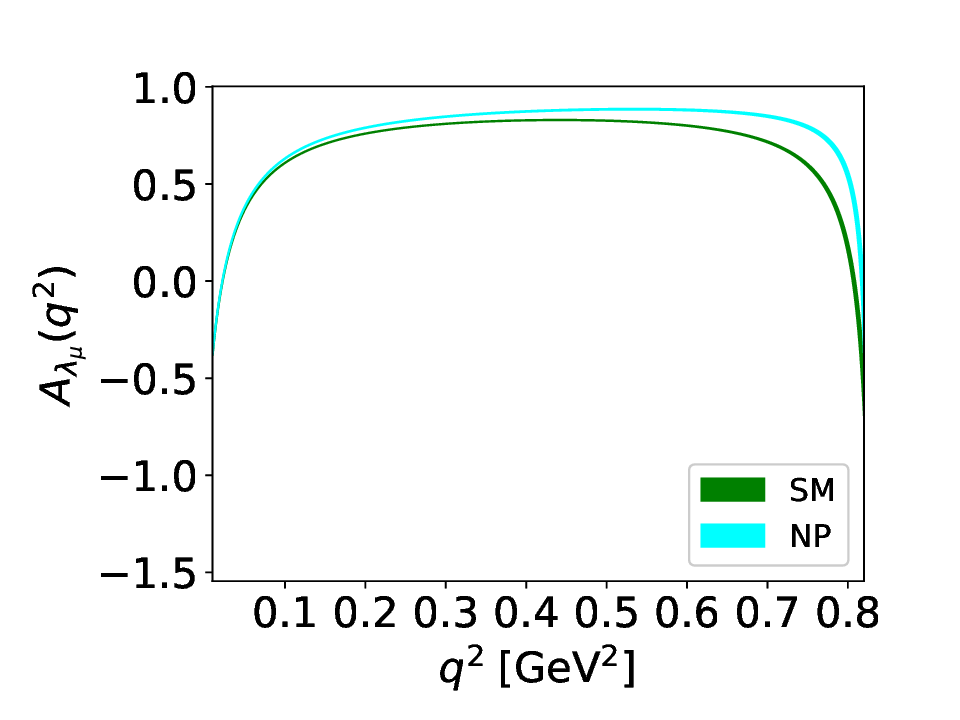}
	\includegraphics[width=0.22\textwidth]{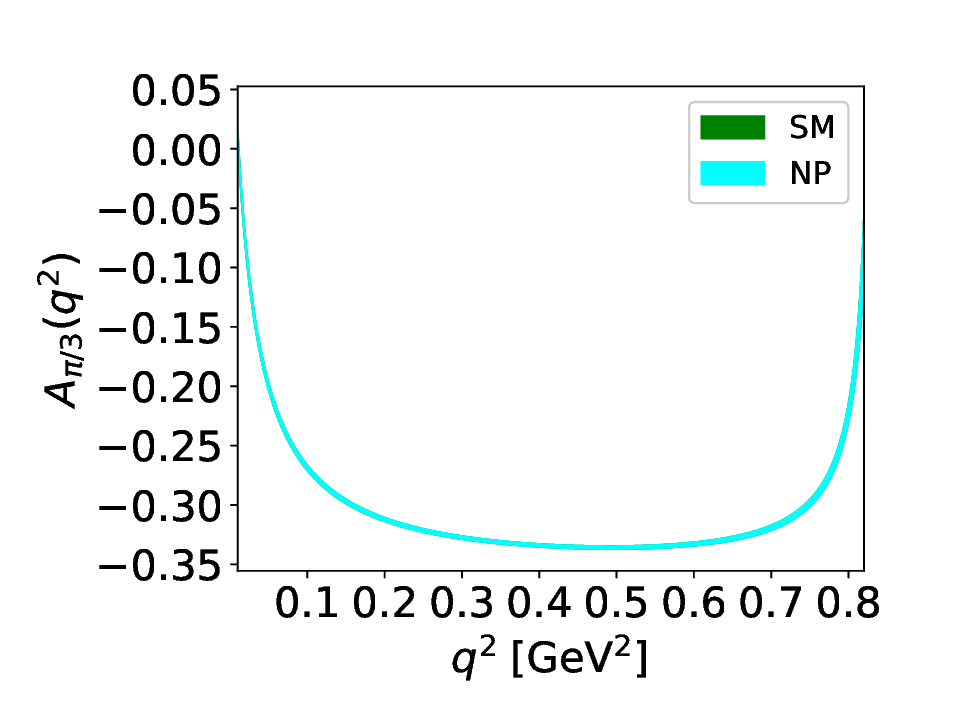}
	\caption{$q^2$ spectra of scenario 8 for the differential branching fraction $\frac{d\mathcal{B}}{dq^2}$ (top left), forward-backward asymmetry $A_{FB}$ (top right), lepton polarization asymmetry $A_{\lambda_{\mu}}$ (bottom left), and convexity $A_{\pi/3}$ (bottom right). The Standard Model (SM green band) and new physics (NP cyan band) both include uncertainty from the form factors.}
	\label{fig-13}       % Give a unique label
\end{figure}

As in figure \ref{fig-13}, the lepton polarization asymmetry gives a small deviation from SM whereas the other three observables for the predictions in semileptonic $B_c^+ \to B_s \mu^+ \nu_{\mu}$ decay don't provide any deviation from the SM.

\section{New Physics Sensitivity Summary}
\label{summary}
In table \ref{tab-senstivity}, we list the sensitivity of all the observables i.e. differential branching fraction, forward-backward asymmetry, lepton polarization asymmetry, and convexity for all the considered scenarios. The sensitive observables for any scenario are green-marked (\greencheck) and insensitive marked by black-cross (\blackcross).

\begin{table}% [H]
	\setlength{\tabcolsep}{4pt}
	\caption{Sensitivity (green tick) and insensitivity (black cross) of different observables for all considered scenarios.}
	\label{tab-senstivity}
	\vspace{0.1cm}
	\centering\begin{tabular}{lll|llll}
		\hline
		\multicolumn{2}{c}{Observables }&$\rightarrow$ &$\frac{d\mathcal{B}}{dq^2}$&$\mathcal{A}_{FB}$ & $\mathcal{A}_{\lambda\mu}$ & $\mathcal{A}_{\pi/3}$\\
		\cline{1-3}
		S.No.& Scenarios &$\downarrow$&&&&\\
		\hline
		1. & Scenario &1& \blackcross & \blackcross &\greencheck & \greencheck \\
		\hline
		2. &Scenario &2& \blackcross & \blackcross & \greencheck & \blackcross\\
		\hline
		3. &Scenario &3& \blackcross & \blackcross &\blackcross & \blackcross\\
		\hline 
		\multirow{2}{*}{4.} & \multirow{2}{*}{Scenario} &4a& \blackcross & \blackcross &\greencheck & \blackcross\\
		%\cline{4-7}
		&&4b& \blackcross & \greencheck &\greencheck & \greencheck\\
		\hline
		\multirow{2}{*}{5.} & \multirow{2}{*}{Scenario}&5a& \blackcross & \blackcross & \greencheck & \blackcross\\
		%\cline{4-7}
		&&5b& \blackcross & \greencheck &\greencheck & \greencheck\\
		\hline
		6. &Scenario &6& \greencheck & \greencheck &\greencheck & \greencheck\\
		\hline
		\multirow{2}{*}{7.} & \multirow{2}{*}{Scenario}&7a& \greencheck & \greencheck &\greencheck & \greencheck\\
		%\cline{4-7}
		&&7b& \greencheck & \greencheck &\greencheck & \greencheck\\
		\hline
		8. &Scenario &8& \blackcross & \blackcross & \greencheck & \blackcross\\
		\hline
	\end{tabular}
\end{table}

\section{Conclusion}
\label{conclusion}
In this study, we investigate the new physics effects in semileptonic decays governed by the quark level transition $c \to s \ell \nu$ with right-handed neutrino. However, the experimental measurement in charm decays is consistent with SM prediction within uncertainties but it can constrain the new physics parameter space.
We have performed a comprehensive analysis of the $B_c^+ \to B_s \mu^+ \nu_{\mu}$ decay for the effects of new physics with a probe of right-handed neutrino. We provide the analytical results for the differential branching fraction, forward-backward asymmetry, lepton polarization asymmetry, and convexity parameter in the decay $B_c^+ \to B_s \mu^+ \nu_{\mu}$ which is governed by the quark level transition $c \to s \ell \nu$. \\
In our numerical analysis, we work in the general effective field theory considering low energy effective Hamiltonian for $c \to s \mu^+ \nu_{\mu}$ with right-handed neutrino. We constrained the allowed parameter space by using the available experimental measurement for the processes governed by $c \to s \mu^+ \nu_{\mu}$ quark level transition.
We have used the form factor for this decay obtained in ref \cite{Cooper:2020wnj} with chained fit by combining the NRQCD and heavy-HISQ results. 
The constrained new physics couplings are used to predict the departure of the different observables in the semileptonic decay $B_c^+ \to B_s \mu^+ \nu_{\mu}$ from SM.
We provide the $q^2$ spectrum prediction of the differential branching fraction, forward-backward asymmetry, lepton polarization asymmetry, and convexity parameter in this decay within the SM and for the constrained NP couplings for different NP scenarios. The central values of NP WCs is considered for each of the NP scenarios for predictions of different observables. It is found that most of the scenarios provide significant deviation from the SM and can be distinguished. The zero crossing shifts from the SM for lepton polarization asymmetry in some NP scenarios which will be interesting to distinguish. The future measurement of this decay at BESIII, and LHCb will guide either validating the SM or uncovering new physics interactions. The current SM and NP predictions for the decay remain limited due to uncertainties in hadronic inputs and higher-order corrections. It is premature to make a definitive statement in terms of the requirement of experimental precision required in order to be able to distinguish between the SM and an NP scenario. After the upgrade of LHCb is expected to achieve an integrated luminosity of $50 fb^{-1}$ by the end of Run 3 so the measurement of this decay can be possible in Run 3 of LHCb.
%%%%%%%%%%%%%%%%%%%%%%%%%%%%%%%%%%%%%%%%%%%%%%%%%%%%%%%%%%%%

\section*{Acknowledgements}
The work of DK is supported by the SERB, Govt. of India under the research grant no. SERB/EEQ/2021/000965. 

%%%%%%%%%%%%%%%%%%%%%%%%%%%%%%%%%%%%%%%%%%%%%%%%%%%%%%%%%%%%%%%%%%%
\appendix{}

%\section{Appendices}

\section{Hadronic Amplitudes}
\label{app-A}
The hadronic matrix elements can be given by showing their 
Lorentz structure as
\begin{eqnarray}
\label{equ-a1}
\begin{split}
& \left\langle B_s\left(p^{\prime}\right)\left|\bar{s} \gamma_\mu c\right| B_c(p)\right\rangle = \\
& f_{+}^{B_c \rightarrow {B_{s}}}\left(q^2\right)\left(p_\mu+p_\mu^{\prime}-\frac{m_{B_c}^2-m_{B_{s}}^2}{q^2} q_\mu\right) \\
& + f_0^{B_c \rightarrow {B_{s}}}\left(q^2\right) \frac{m_{B_c}^2-m_{B_{s}}^2}{q^2} q_\mu, \\
& \left\langle {B_{s}}\left(p^{\prime}\right)|\bar{s} c| B_c(p)\right\rangle = f_S^{B_c \rightarrow {B_{s}}}\left(q^2\right) = \\ & \frac{m_{B_c}^2-m_{B_{s}}^2}{m_c-m_s} f_0^{B_c \rightarrow {B_{s}}}\left(q^2\right), \\
& \left\langle {B_{s}}\left(p^{\prime}\right)\left|\bar{s} \sigma_{\mu \nu} c\right| B_c(p)\right\rangle = \\
& -i \frac{2 f_T^{B_c \rightarrow {B_{s}}}\left(q^2\right)}{m_{B_c}+m_{B_{s}}}\left(p_\mu p_\nu^{\prime}-p_\nu p_\mu^{\prime}\right), \\
& \left\langle {B_{s}}\left(p^{\prime}\right)\left|\bar{s} \sigma_{\mu \nu} \gamma_5 c\right| B_c(p)\right\rangle = -\frac{2 f_T^{B_c \rightarrow {B_{s}}}\left(q^2\right)}{m_{B_c}+m_{B_{s}}} \epsilon_{\mu \nu \alpha \beta} p^\alpha p^{\prime \beta}.
\end{split}
\end{eqnarray}
with $\epsilon^{0123} = +1$.

The hadronic helicity amplitudes are given as
\begin{eqnarray}
\label{equ-a3}
\begin{aligned}
&	H_{V, 0}^s(q^2) = \sqrt{\frac{\lambda_{B_{s}}(q^2)}{q^2}} f_+(q^2), \\
&	H_{V, t}^s(q^2) = \frac{m_{B_{c}}^2-m_{B_{s}}^2}{\sqrt{q^2}} f_0(q^2), \\
&	H_S^s(q^2) = \frac{m_{B_{c}}^2-m_{B_{s}}^2}{m_c-m_s} f_0(q^2), \\
&	H_T^s(q^2) = -\frac{\sqrt{\lambda_{B_{s}}(q^2)}}{m_{B_{c}}+m_{B_{s}}} f_T(q^2),
\end{aligned}
\end{eqnarray}

 \section{Form factors and helicity amplitudes for $D \to P$ and $D \to V$ transitions}
 \label{app-B}
\subsection{$D \to P$}
The hadronic helicity amplitudes following from \cite{Sakaki:2013bfa} are given by
\begin{eqnarray}
\label{equ-b1}
\begin{aligned}
&H_{V, 0}^P(q^2) =\sqrt{\frac{\lambda_P(q^2)}{q^2}} f_+(q^2), \\  &H_{V, s}^P(q^2) =\frac{m_{D}^2-m_{P}^2}{\sqrt{q^2}} f_0(q^2), \\
&H_{S,s}^P(q^2) =\frac{m_{D}^2-m_{P}^2}{m_c-m_s} f_0(q^2), \\ &H_T^P(q^2) =-\frac{\sqrt{\lambda_P(q^2)}}{m_{D}+m_{P}} f_T(q^2),\\
&H_{P, s}^P(q^2) = 0, \quad H_{V, \pm}^P(q^2) = 0 
\end{aligned}	
\end{eqnarray}
For form factor in our work, we use the z-series parameterization ($z_0 = z(0,q^2)$) from 
Ref.\cite{Lubicz:2017syv, Lubicz:2018rfs} (Table \ref{tab-4})
\begin{equation}
f_{+}(q^2) = \frac{f(0) + c_{+}(z-z_{0})(1 + \frac{1}{2}(z+z_{0}))}{1 - \frac{q^2}{M^2_{D_s^*}}}
\end{equation}
\begin{equation}
f_{0}(q^2) = f(0) + c_{0}(z-z_{0})(1 + \frac{1}{2}(z+z_{0}))
\end{equation}
\begin{equation}
f_{T}(q^2) = \frac{f_T(0) + c_{T}(z-z_{0})(1 + \frac{1}{2}(z+z_{0}))}{1 - P_{T}q^2}
\end{equation}

\begin{table}%[H]
	\begin{center}
		\setlength{\tabcolsep}{30pt}
		\caption{Fit parameters for the $f_{+}, f_{0}$ and $f_{T}$ in the z-series expansion.}
		\label{tab-4}       % Give a unique label
		\begin{tabular}{ll}
			\hline\noalign{\smallskip}
			Fit Parameters & Values   \\
			\noalign{\smallskip}\hline\noalign{\smallskip}
			$f_{0}$ & 0.7647 (308) \\ 
			$c_{+}$ & -0.066 (333) \\
			$c_{0}$ & -2.084 (283) \\
			$f_{T}(0)$ & 0.6871 (542) \\
			$c_{T}$ & -2.86 (1.46) \\
			$P_{T} (GeV)^{-2}$  & 0.0854 (671) \\
			\noalign{\smallskip}\hline
		\end{tabular}
	\end{center}
\end{table}

\subsection{$D \to V$}
The Hadronic helicity amplitude for $D \to V$ decays is defined as follows \cite{Sakaki:2013bfa}.
\begin{eqnarray}
\begin{aligned}
&H_{V, \pm}^V = (m_D + m_V)A_1(q^2) \mp \frac{\sqrt{\lambda_V(q^2)}}{(m_D + m_V)}V(q^2)\\
&H_{V, 0}^V = \frac{(m_D + m_V)}{2 m_V \sqrt{q^2}}[-(m_D^2 - m_V^2 - q^2)A_1(q^2) \\
& + \frac{\sqrt{\lambda_V(q^2)}}{(m_D + m_V)^2}A_2(q^2)], \\
&H_{V,t}^V =\sqrt{\frac{\lambda_V(q^2)}{q^2}} A_0(q^2), \\
&H_P^V =-\frac{\sqrt{\lambda_V(q^2)}}{m_{c}+m_{s}} A_0(q^2),\\
&H_{T, \pm}^V = \frac{1}{\sqrt{q^2}} \Big[\pm (m_D^2 - m_V^2)T_2(q^2) + \sqrt{\lambda_V(q^2)}T_1(q^2)\Big], \\
&H_{T, 0}^V = \frac{1}{2m_V} \Big[- (m_D^2 + 3m_V^2 - q^2)T_2(q^2) \\ &+ \frac{\lambda_V(q^2)}{m_D^2 - m_V^2}T_3(q^2)\Big]
\end{aligned}
\end{eqnarray}

In our work, we utilize the form factor of \cite{Fleischer:2019wlx, Wu:2006rd, bowler1995improved} extracted by following the lattice double-pole parameterization.
\begin{eqnarray}
\begin{split}
&A_i(q^2) = \frac{A_i(0)}{1 - \frac{a_i q^2}{m_D^2} + b_i\big(\frac{q^2}{m_{D^2}}\big)^2} \\
&V(q^2) = \frac{V(0)}{1 - \frac{a_V q^2}{m_D^2} + b_V\big(\frac{q^2}{m_{D^2}}\big)^2}
\end{split}   
\end{eqnarray}

where $\mathrm{i}$ = 1,2,3 for $A_1, A_2, A_3$ form factors respectively.

\begin{table}[H]
	\begin{center}
		\setlength{\tabcolsep}{7pt}
		\caption{Fit parameters for the form factors obtained by considering the leading twist meson distribution amplitudes \cite{Wu:2006rd}.}
		\label{tab-5}       % Give a unique label
		\begin{tabular}{lll}
			\hline\noalign{\smallskip}
			\bf{Decay Mode $\to$}&$ D \to K^*$&$ D_s^+ \to \phi$ \\
			\noalign{\smallskip}\hline\noalign{\smallskip}
			Fit parameters & Values   & Values \\
			\noalign{\smallskip}\hline\noalign{\smallskip}
			$A_1(0)$ & 0.601 $\pm$ 0.030 & 0.605 $\pm$ 0.043 \\ 
			$a_1$ & 0.51 $\pm$ 0.02 & 0.48 $\pm$ 0.02 \\
			$b_1$ & 0.04 $\pm$ 0.01 & -0.007 $\pm$ 0.003 \\
			$A_2(0)$ & 0.541 $\pm$ 0.038 & 0.583 $\pm$ 0.038 \\
			$a_2$ & 0.91 $\pm$ 0.10 & 0.70 $\pm$ 0.01 \\
			$b_2$ & -0.68 $\pm$ 0.21 & 0.16 $\pm$ 0.03 \\
			$A_3(0)$ & -0.541 $\pm$ 0.038&-0.583 $\pm$ 0.038 \\
			$a_3$ & 0.91 $\pm$ 0.10&0.70 $\pm$ 0.01 \\
			$b_3$ & -0.68 $\pm$ 0.21 & 0.16 $\pm$ 0.03 \\
			$V(0)$ & 0.796 $\pm$ 0.032 & 0.951 $\pm$ 0.053 \\
			$a_V$ & 0.60 $\pm$ 0.13 & 0.98 $\pm$ 0.05 \\
			$b_V$ & 1.53 $\pm$ 0.30 & 0.57 $\pm$ 0.02 \\
			\noalign{\smallskip}\hline
		\end{tabular}
	\end{center}
\end{table}

And relations to relate the penguin-type form factors with the semileptonic type ones are given as
\vspace{-3mm}
\begin{equation}
\begin{split}
& A_0(q^2) = 
\frac{1}{2 m_V} \Big((m_D + m_V) A_1(q^2) -  (m_D + m_V) A_2(q^2) \\
& - \frac{q^2}{m_D + m_V} A_3(q^2) \Big) \\
& T_1(q^2) = 
\Big(\frac{m_D^2 - m_V^2 + q^2}{2 m_D} \Big) \Big(\frac{V(q^2)}{m_D + m_V} \Big) \\
& + \Big(\frac{m_D + m_V}{2 m_D}\Big) A_1(q^2) \\
&T_2(q^2) =
\frac{2}{m_D^2 - m_V^2} \Big[\frac{(m_D - y)(m_D + m_V)}{2} A_1(q^2) \\
& + \frac{m_D(y^2 - m_V^2)}{m_D + m_V} V(q^2)\Big] \\
&T_3(q^2) =
-\frac{m_D + m_V}{2 m_D} A_1(q^2) \\ 
& + \frac{m_D - m_V}{2 m_D} \big[ A_2(q^2) - A_3(q^2)\big]
 + \frac{m_D^2 + 3 m_V^2 - q^2}{2 m_D(m_D + m_V)} V(q^2)
\end{split}
\end{equation}
where $y = \frac{m_D^2 + m_V^2 - q^2}{2 m_D}$

\bibliography{reference}

\end{document}